\documentclass[twocolappendix]{aastex631}

\hypersetup{linkcolor=red,citecolor=blue,filecolor=cyan,urlcolor=magenta}

\def\lessim{\mathrel{\hbox{\rlap{\hbox{\lower4pt\hbox{$\sim$}}}\hbox{$<$}}}}
\def\grtsim{\mathrel{\hbox{\rlap{\hbox{\lower4pt\hbox{$\sim$}}}\hbox{$>$}}}}

\usepackage{CJK}

\shorttitle{M31 MMRD}
\shortauthors{Clark et al.}

\graphicspath{{./}{figures/}}

\begin{document}
\title{Exploring the MMRD Relation for Novae in M31}

\correspondingauthor{A. W. Shafter}
\email{ashafter@sdsu.edu}

\author[0000-0003-4735-9128]{J. Grace Clark}
\affiliation{Department of Astronomy, San Diego State University, San Diego, CA 92182, USA}

\author[0000-0002-0835-225X]{Kamil Hornoch}
\affiliation{Astronomical Institute of the Czech Academy of Sciences, Fri\v{c}ova 298, CZ-251 65 Ond\v{r}ejov, Czech Republic}

\author[0000-0002-1276-1486]{Allen W. Shafter}
\affiliation{Department of Astronomy, San Diego State University, San Diego, CA 92182, USA}

\author[0000-0002-1330-1318]{Hana Ku\v{c}\'akov\'a}
\affiliation{
Astronomical Institute, Charles University, Faculty of Mathematics and Physics, V Hole\v{s}ovi\v{c}k\'ach 2,
CZ-180 00 Prague, Czech Republic}
\affiliation{Research Centre for Theoretical Physics and Astrophysics, Institute of Physics, Silesian University
in Opava, Bezru\v{c}ovo n\'am. 13, CZ-746 01 Opava, Czech Republic}

\author{Jan Vra\v{s}til}
\affiliation{
Astronomical Institute, Charles University, Faculty of Mathematics and Physics, V Hole\v{s}ovi\v{c}k\'ach 2,
CZ-180 00 Prague, Czech Republic}

\author[0000-0001-6098-6893]{Peter Ku\v{s}nir\'ak}
\affiliation{Astronomical Institute of the Czech Academy of Sciences, Fri\v{c}ova 298, CZ-251 65 Ond\v{r}ejov, Czech Republic}

\author[0000-0002-4387-6358]{Marek Wolf}
\affiliation{
Astronomical Institute, Charles University, Faculty of Mathematics and Physics, V Hole\v{s}ovi\v{c}k\'ach 2,
CZ-180 00 Prague, Czech Republic}

\begin{abstract}

The results of a two decade long $R$-band photometric survey of novae in M31 are presented. From these data,
$R$-band light curves have been determined for 180 novae with data sufficient for estimating peak
brightness and subsequent rate of decline.
The data show a weak correlation of peak brightness
with fade rate consistent with the
well-known Maximum Magnitude versus Rate of Decline (MMRD) relation. As generally appreciated for Galactic novae, the large scatter in the MMRD
relation precludes its use in determining distances to individual novae.
The novae at maximum light are distributed with standard deviation $\sigma=0.89$~mag about a mean $R$-band absolute magnitude
given by $\langle M_R \rangle=-7.57\pm0.07$.
The overall M31 luminosity distribution
is in excellent agreement with that found for Galactic novae suggesting that the nova populations in M31
and the Galaxy are quite similar.
The notion that all novae can be characterized by a standard luminosity 15~d after maximum light ($M_{15}$) is also explored. Surprisingly,
the distribution of $M_{15}$ values is characterized by a standard deviation only slightly smaller than that for novae at maximum light and thus
offers little promise for precise extragalactic distance determinations.
A dozen faint and fast novae that are likely to be previously unidentified recurrent novae
have been identified from their position in the MMRD plot and in the $M_{15}$ distribution.

\end{abstract}


\section{Introduction} \label{sec:intro}

Classical Novae are all close binary systems where a white dwarf primary star accretes mass from
a late-type companion \citep[e.g., see][]{1995cvs..book.....W}. The transferred material slowly becomes degenerate
and once the temperature and density at the base of the accreted envelope become sufficiently
high to initiate hydrogen burning a thermonuclear runaway (TNR) ensues resulting in a nova eruption that drives significant mass loss from
the system \citep[e.g., see][and references therein]{2016PASP..128e1001S,2008clno.book.....B}. At the peak of their eruptions novae can reach
absolute visual magnitudes as high as $M_V\sim-10$ making them among the brightest astrophysical transients known, and
easily visible in nearby galaxies with moderate-sized telescopes.

It has long been appreciated that novae exhibit a wide range of peak luminosities and subsequent rates of decline. Initial attempts at exploring
correlations between these properties were confounded by the failure to treat novae and supernovae as distinct phenomena.
This confusion led \citet{1936PASP...48..191Z} to propose a ``life-luminosity" relation for all novae where the lifetime (given by the time in days to fade by two magnitudes from maximum light, $t_2$) was
positively correlated with the peak luminosity. In other words, the brighter systems (the supernovae) evolved the slowest (i.e., had longer $t_2$). After supernovae were eventually understood as a separate phenomena and omitted from consideration, it was finally realized that the ``common novae" as they were referred to at the time, behaved in the opposite fashion. In particular, in his study of a small sample of common novae \citet{1942PA.....50..233M} concluded
that ``Since Nova Aquilae and Nova Persei, whose luminosities were of the highest, were both very fast novae, there is some suggestion of a
connection between luminosity and rate of decline, or a `life-luminosity relation'. But there is no reason to believe that it is more than a statistical trend; a precise correlation is not implied." However, within three years
\citet{1945PASP...57...69M} had strengthened his case, arguing that the peak luminosity of a classical nova was indeed
correlated with its rate of decline from maximum light with
the most luminous novae experiencing the most rapid decline from peak luminosity.
Over the years, the ``life-luminosity" relation of McLaughlin has come to be known as the Maximum-Magnitude versus Rate-of-Decline (MMRD) relation, and it has been calibrated many times, both in the Galaxy and in M31 \citep[e.g.,][]{1955Obs....75..170B,1956AJ.....61...15A,1960stat.book..585M,1976A&A....50..113P,1985ApJ...292...90C,1989AJ.....97.1622C,2000AJ....120.2007D,2011ApJ...734...12S,2018MNRAS.476.4162O,2022MNRAS.517.6150S}.

Given that novae are significantly brighter than Cepheid variables on average, for the first several decades after the MMRD relation was first proposed there was considerable interest in employing the relation for the determination of distances to nearby
galaxies \citep[e.g.,][]{1955Obs....75..170B,1978ApJ...223..351D,1986PASP...98..110V,1987ApJ...318..507P,1988ASPC....4..221V}.
The initial enthusiasm was soon tempered, however, when it became apparent that the considerable
scatter in the MMRD relation was not simply due to observational errors, but appeared to be
intrinsic to the relation itself. Unlike the period-luminosity relation for Cepheids, the MMRD
seemed at best an approximate correlation.

A favorite target for the study of extragalactic novae has been the Andromeda galaxy (M31) where
almost 1300 nova candidates have been discovered over the past century \citep[e.g., see][and references therein]{2019enhp.book.....S, 2020A&ARv..28....3D}. Although some novae have been discovered serendipitously as part of unrelated studies, the majority have been identified in a handful of long-term targeted surveys \citep[e.g.][]{1929ApJ....69..103H,1956AJ.....61...15A,1964AnAp...27..498R,1973Ros,1987ApJ...318..520C,2001ApJ...563..749S,2022ApJ...936..117R}.
In this paper we discuss the properties of a large
and homogeneous sample of $R$-band nova light curves obtained as part of a long-term survey of the bulge region of M31
conducted primarily at the Ond\v{r}ejov observatory in the Czech Republic. In total, $R$-band data spanning two decades of observation have enabled the photometric evolution of a total of 180 novae to be determined.
The resulting light curves are presented and
analyzed in order to explore both the character of the $R$-band MMRD relation as well as the luminosity distributions at maximum light and at 15 days post maximum, where novae might be expected to reach a similar luminosity regardless of their
peak magnitudes or rates of decline.

\section{Observations}

The majority of observations presented here have been obtained using the 0.65-m reflector at the Ond\v{r}ejov Observatory
(operated partly by the Charles University, Prague) and the 0.35-m telescope in the private observatory of K.H. at
Lelekovice. For complementary observations a variety of additional telescopes (see facilities)
were also used. Standard reduction procedures were applied to the raw CCD images (bias and dark-frame subtraction and flat field correction) using the SIMS\footnote{\tt https://www.gxccd.com/} and Munipack\footnote{\tt https://munipack.physics.muni.cz/} programs. The total exposure time of each series of images varied from a few minutes up to about an hour, with reduced images of the same series being co-added to improve the signal-to-noise ratio. To facilitate nova detection, the gradient of the galaxy background was flattened by the spatial median filter using SIMS. Photometric and astrometric measurements of the novae were then performed using "Optimal Photometry" (based on fitting of point-spread function profiles) in GAIA\footnote{\tt http://star-www.dur.ac.uk/\(\sim \)pdraper/gaia/gaia.html} and APHOT (a synthetic aperture photometry and astrometry software package developed by M. Velen and P. Pravec at the Ondřejov Observatory \citep[e.g., see][]{1994ExA.....5..375P}, respectively. The nova photometry was calibrated using comparison stars in the M31 field from \citet{2006AJ....131.2478M}.
Overall, our sample includes 180 novae for which the temporal coverage was
sufficient to characterize the $R$-band light curves\footnote{A very few measurements were conducted using a Sloan $r$ filter; but due to
the limited number of these observations and the similarity between $r$ and $R$ magnitudes, no corrections were applied in our analysis}. The photometry used in the construction of the light curves is given in the
Data Behind the Figure file.

The spatial coverage of our nova survey covers $\Delta R.A. \simeq 34'$ and $\Delta Decl. \simeq 26'$ centered on the nucleus of M31 -- a region which is heavily dominated by the galaxy's bulge population. Thus, the analysis to follow most closely traces the properties of novae arising in the bulge of M31, although given the high inclination of the plane of M31 to our line of sight, $i\simeq77^{\circ}$ \citep[][]{1978A&A....67...73S}, we cannot exclude the possibility
that some novae from the disk population are included among the novae in our sample.

\section{The Light Curves} \label{sec:LC}

\begin{figure}
\includegraphics[angle=0,scale=0.76]{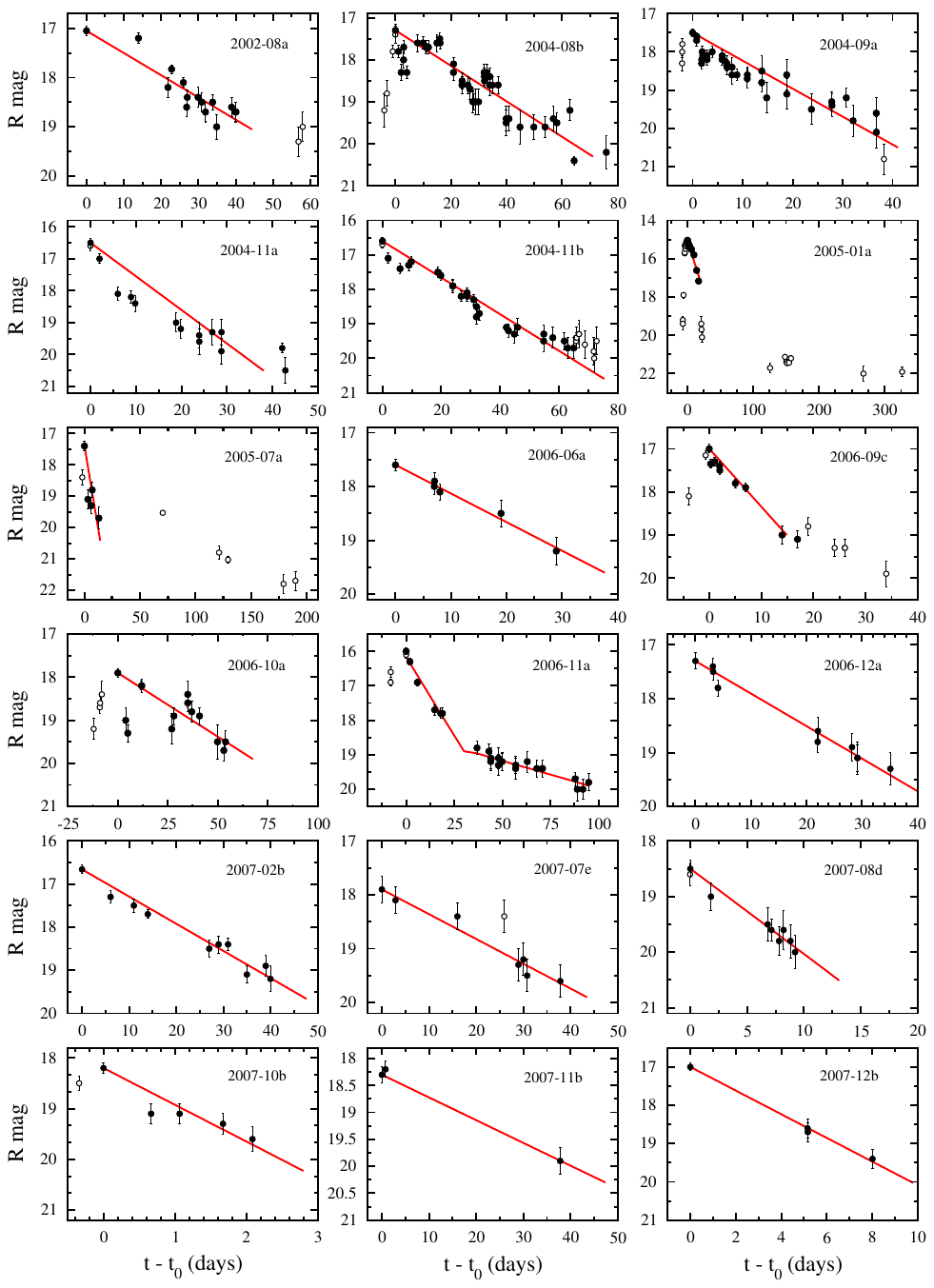}
\caption{The $R$-band light curves for the 180 novae in our M31 sample. The red lines show the best fitting linear declines from peak brightness used to determine the fade rates and $t_2$ times. The open circles show points that were excluded from the fits. The light curve data are presented in Table A1.
}
\label{fig:f1.1}
\end{figure}

\begin{figure}
\renewcommand{\thefigure}{1.2}
\includegraphics[angle=0,scale=0.78]{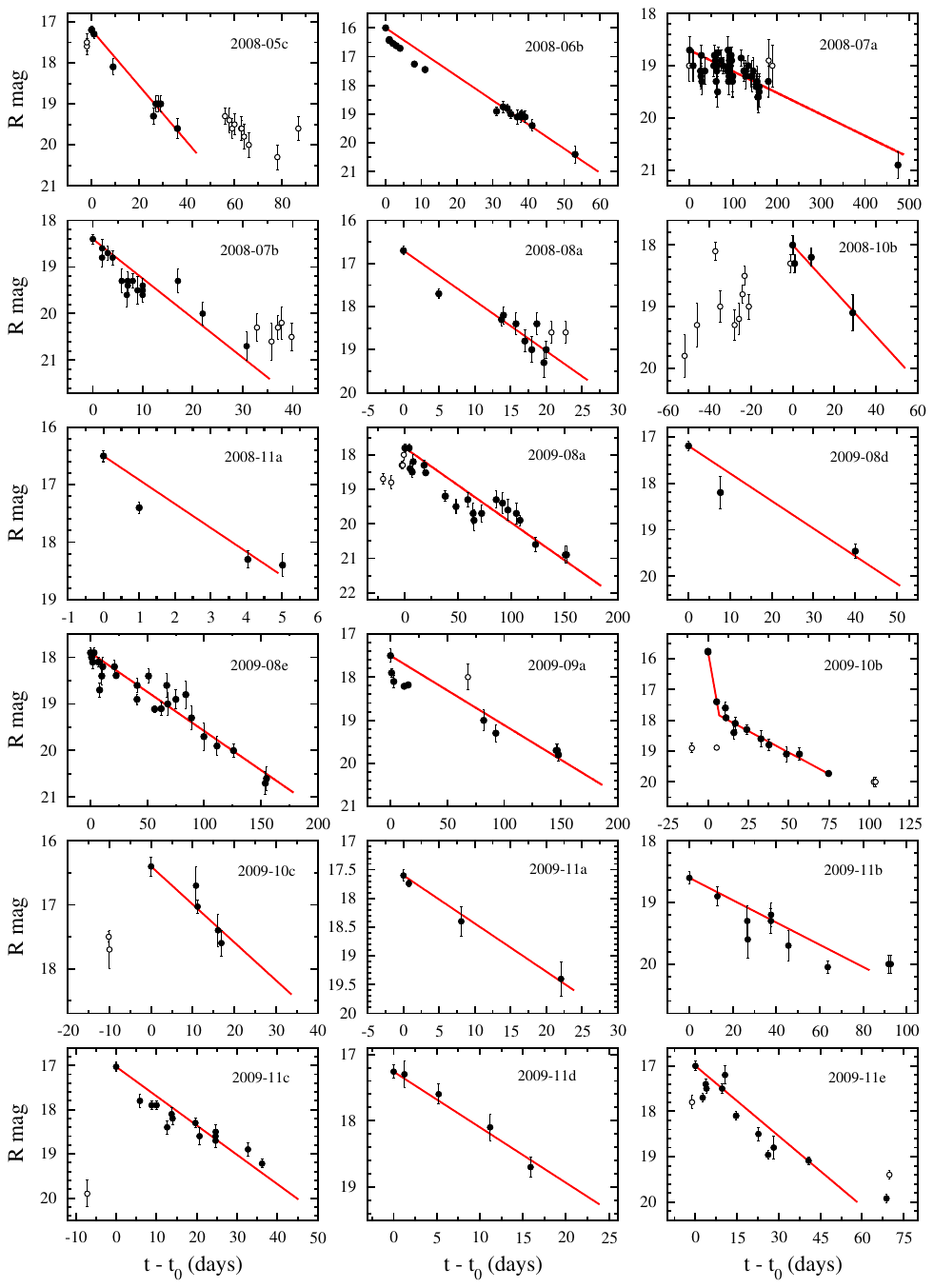}
\caption{Continuation of Figure 1.
}
\label{fig:f1.2}
\end{figure}

\begin{figure}
\renewcommand{\thefigure}{1.3}
\includegraphics[angle=0,scale=0.78]{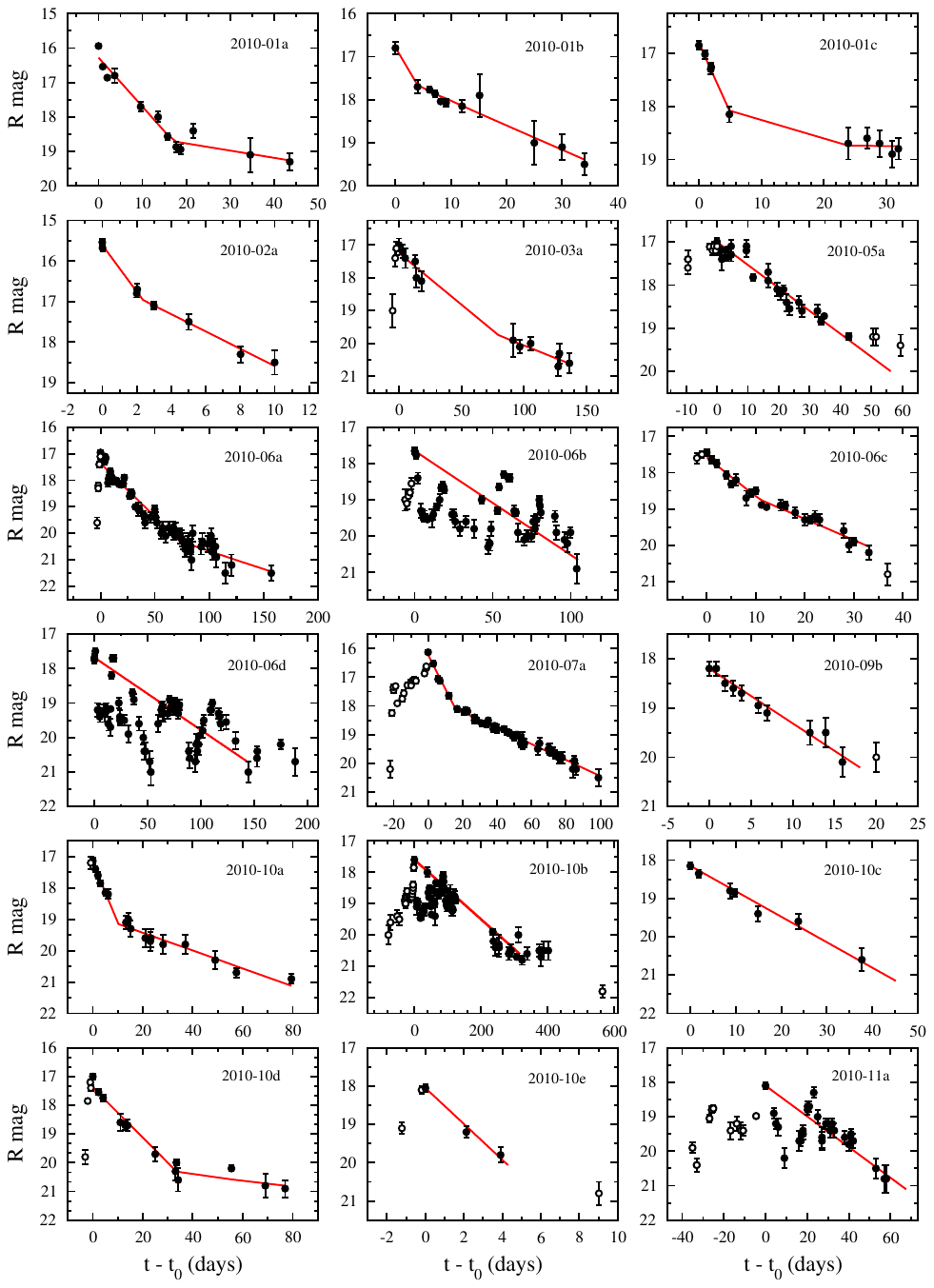}
\caption{Continuation of Figure 1.
}
\label{fig:f1.3}
\end{figure}

\begin{figure}
\renewcommand{\thefigure}{1.4}
\includegraphics[angle=0,scale=0.78]{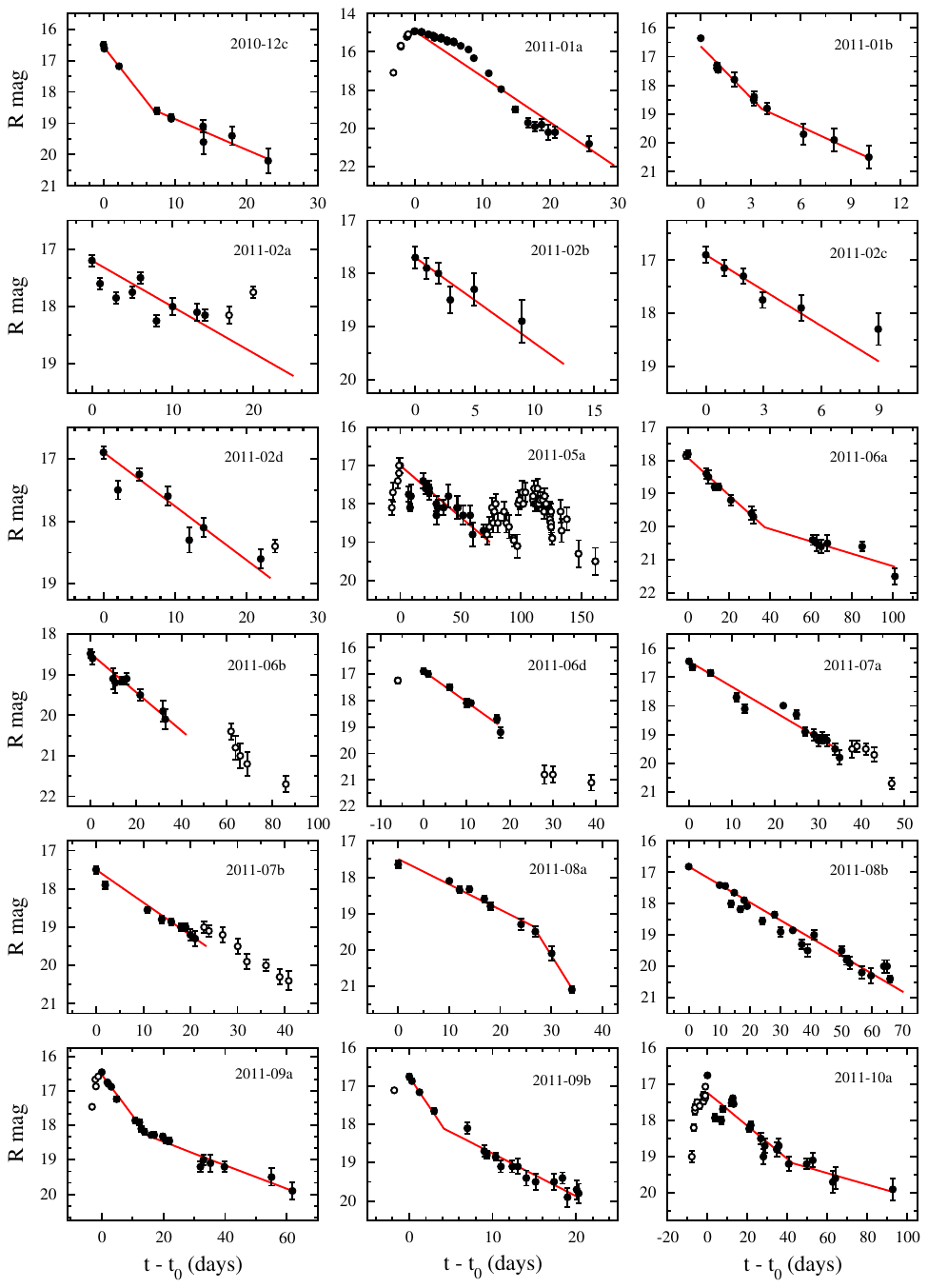}
\caption{Continuation of Figure 1.
}
\label{fig:f1.4}
\end{figure}

\begin{figure}
\renewcommand{\thefigure}{1.5}
\includegraphics[angle=0,scale=0.78]{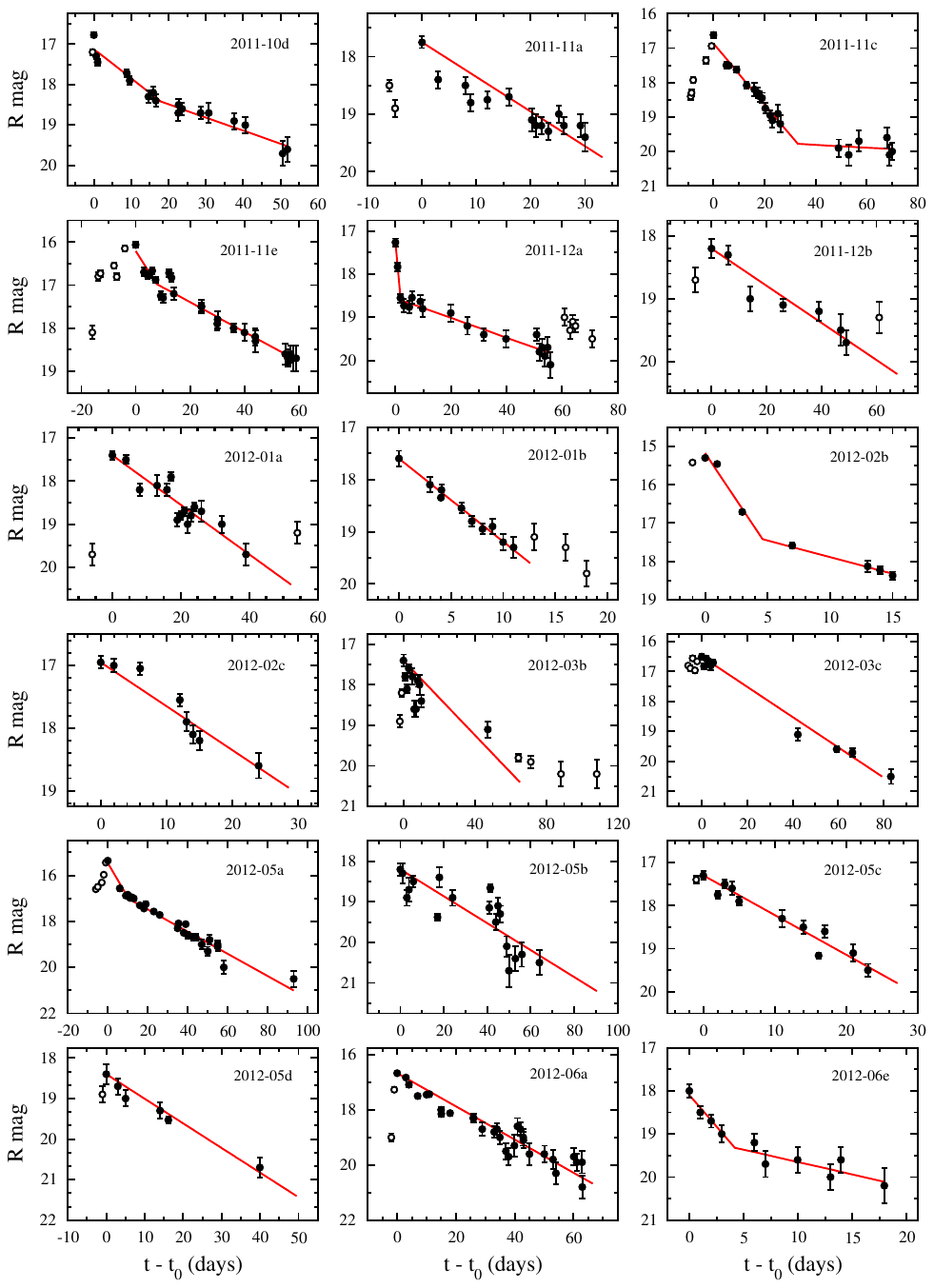}
\caption{Continuation of Figure 1.
}
\label{fig:f1.5}
\end{figure}

\begin{figure}
\renewcommand{\thefigure}{1.6}
\includegraphics[angle=0,scale=0.78]{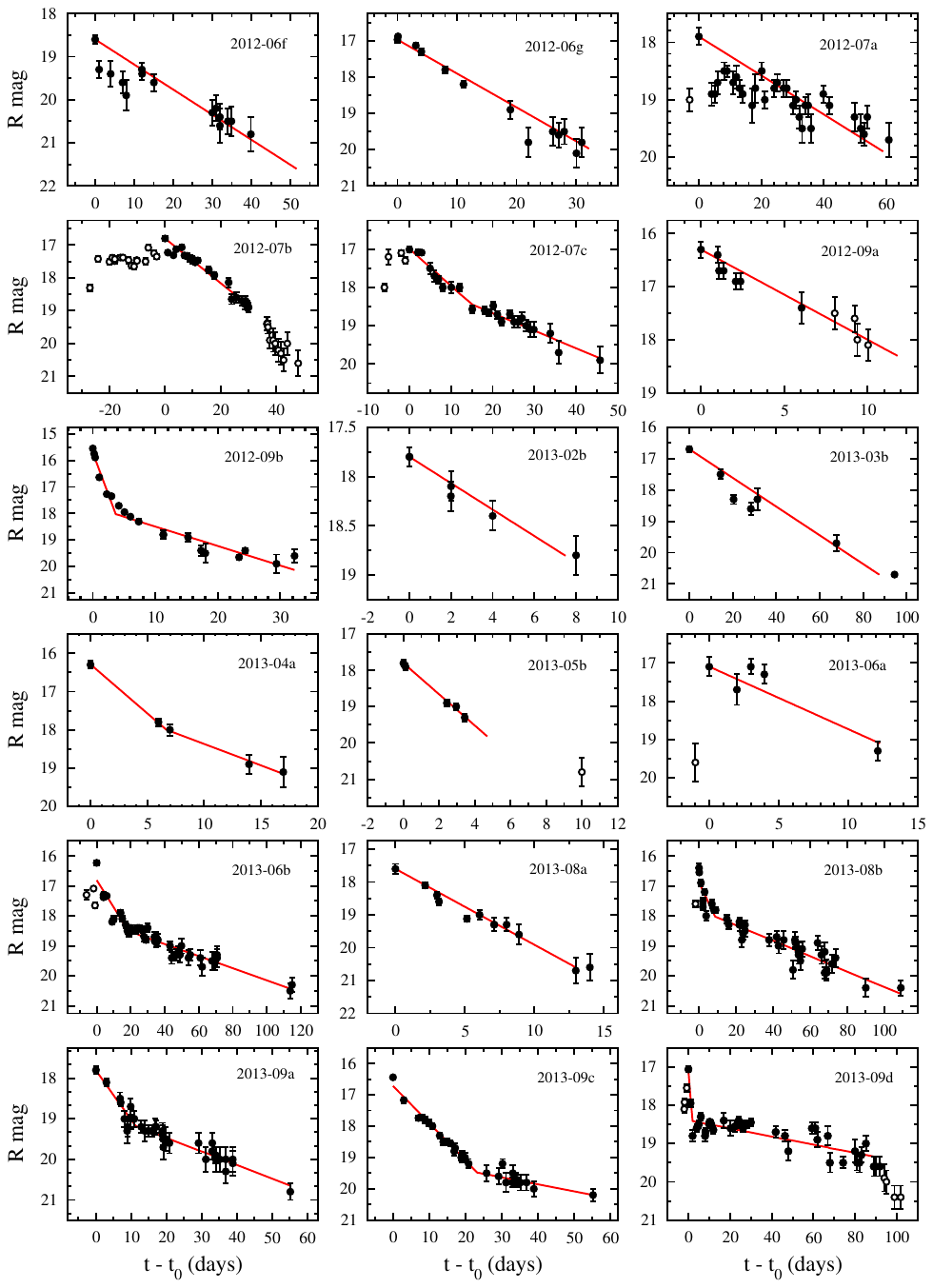}
\caption{Continuation of Figure 1.
}
\label{fig:f1.6}
\end{figure}

\begin{figure}
\renewcommand{\thefigure}{1.7}
\includegraphics[angle=0,scale=0.78]{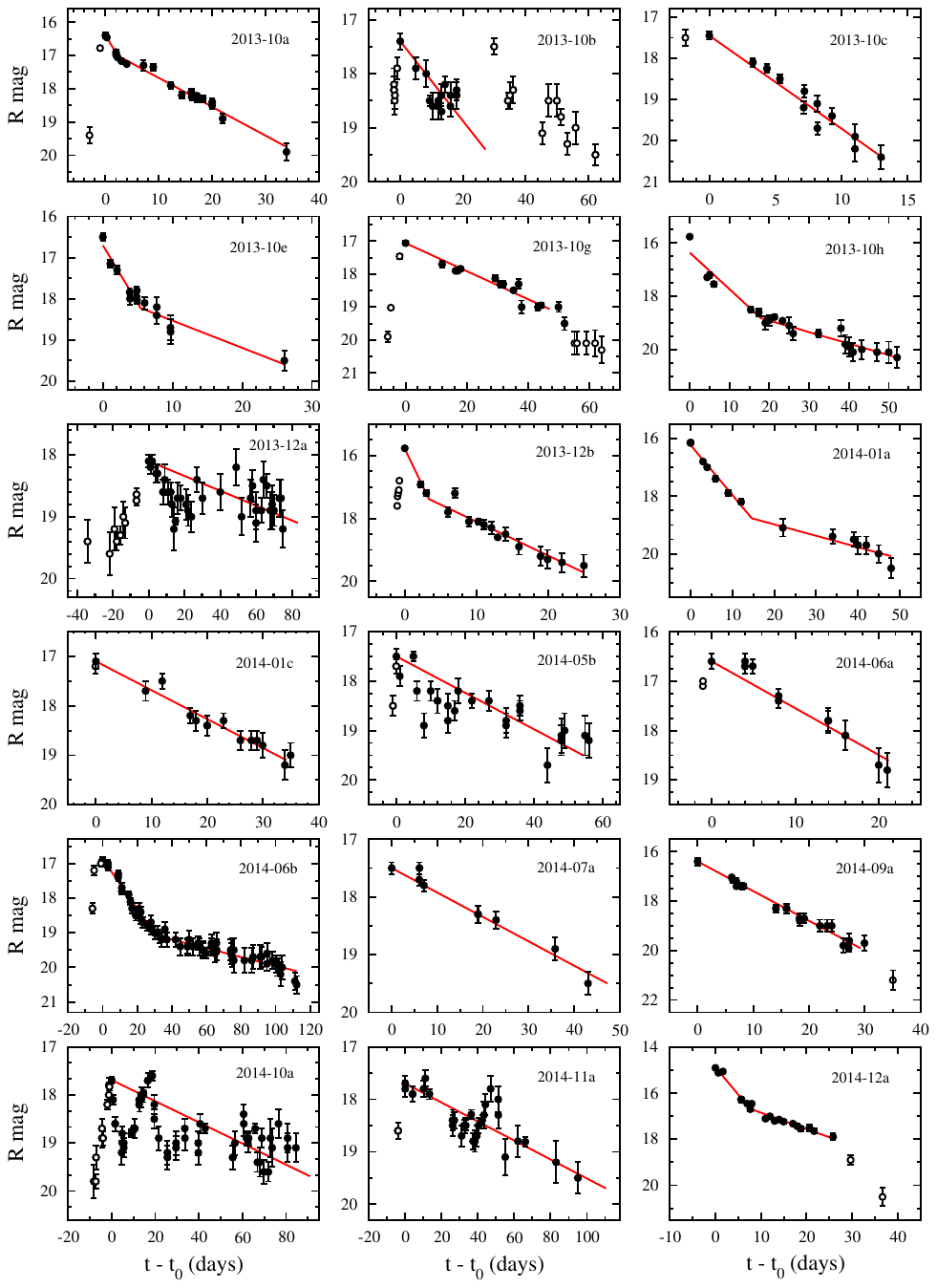}
\caption{Continuation of Figure 1.
}
\label{fig:f1.7}
\end{figure}

\begin{figure}
\renewcommand{\thefigure}{1.8}
\includegraphics[angle=0,scale=0.78]{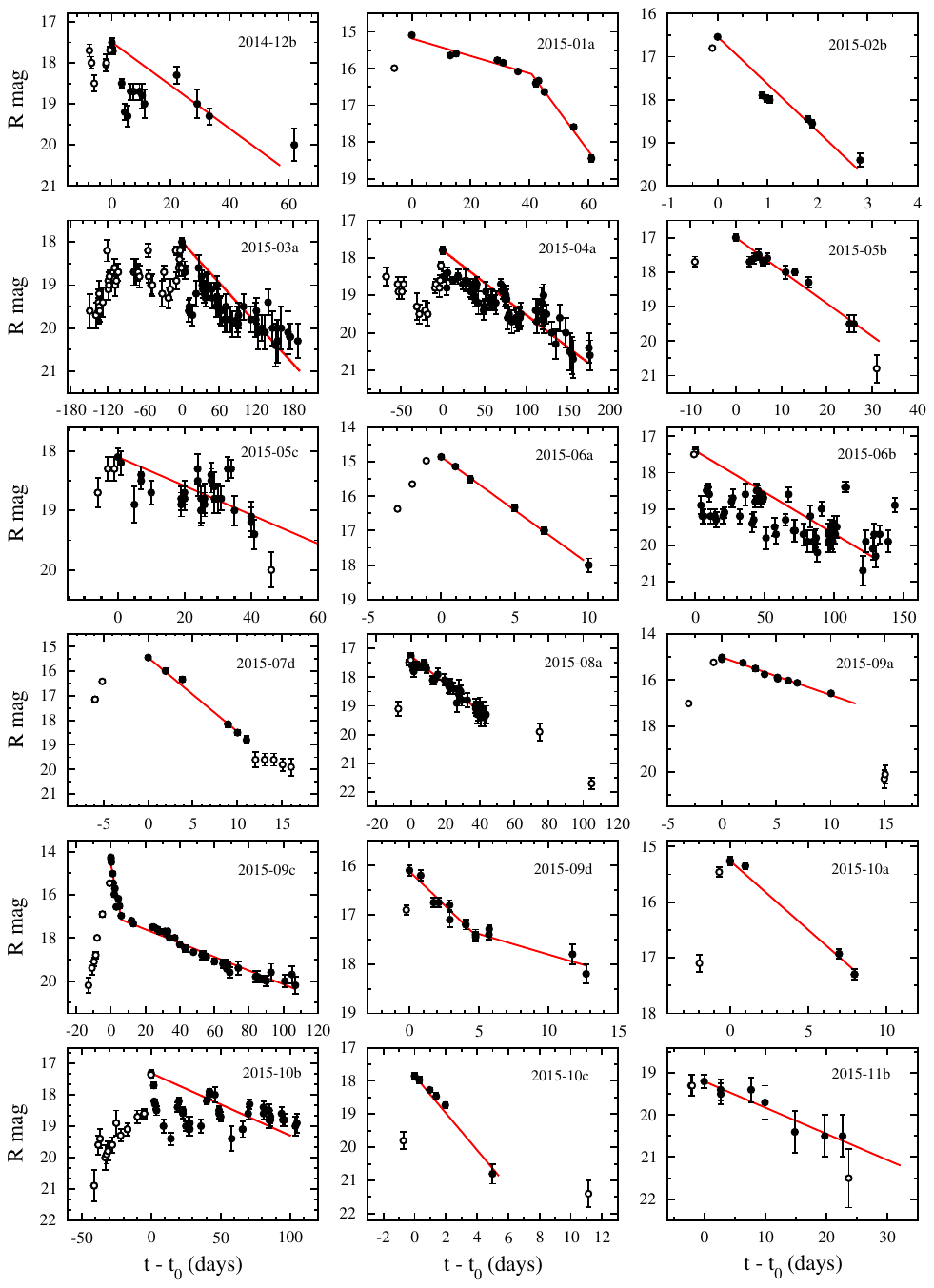}
\caption{Continuation of Figure 1.
}
\label{fig:f1.8}
\end{figure}

\begin{figure}
\renewcommand{\thefigure}{1.9}
\includegraphics[angle=0,scale=0.78]{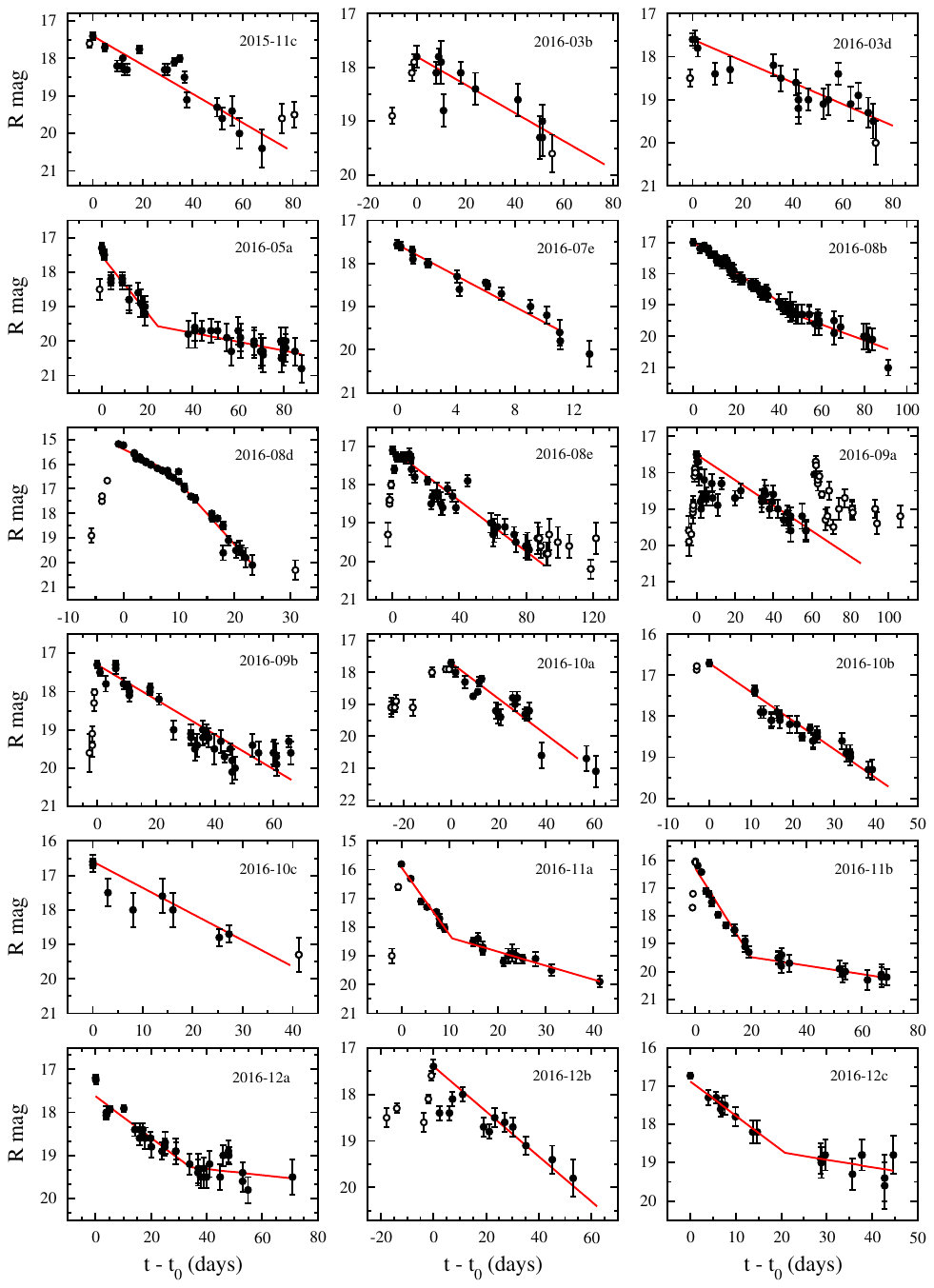}
\caption{Continuation of Figure 1.
}
\label{fig:f1.9}
\end{figure}

\begin{figure}
\renewcommand{\thefigure}{1.10}
\includegraphics[angle=0,scale=0.78]{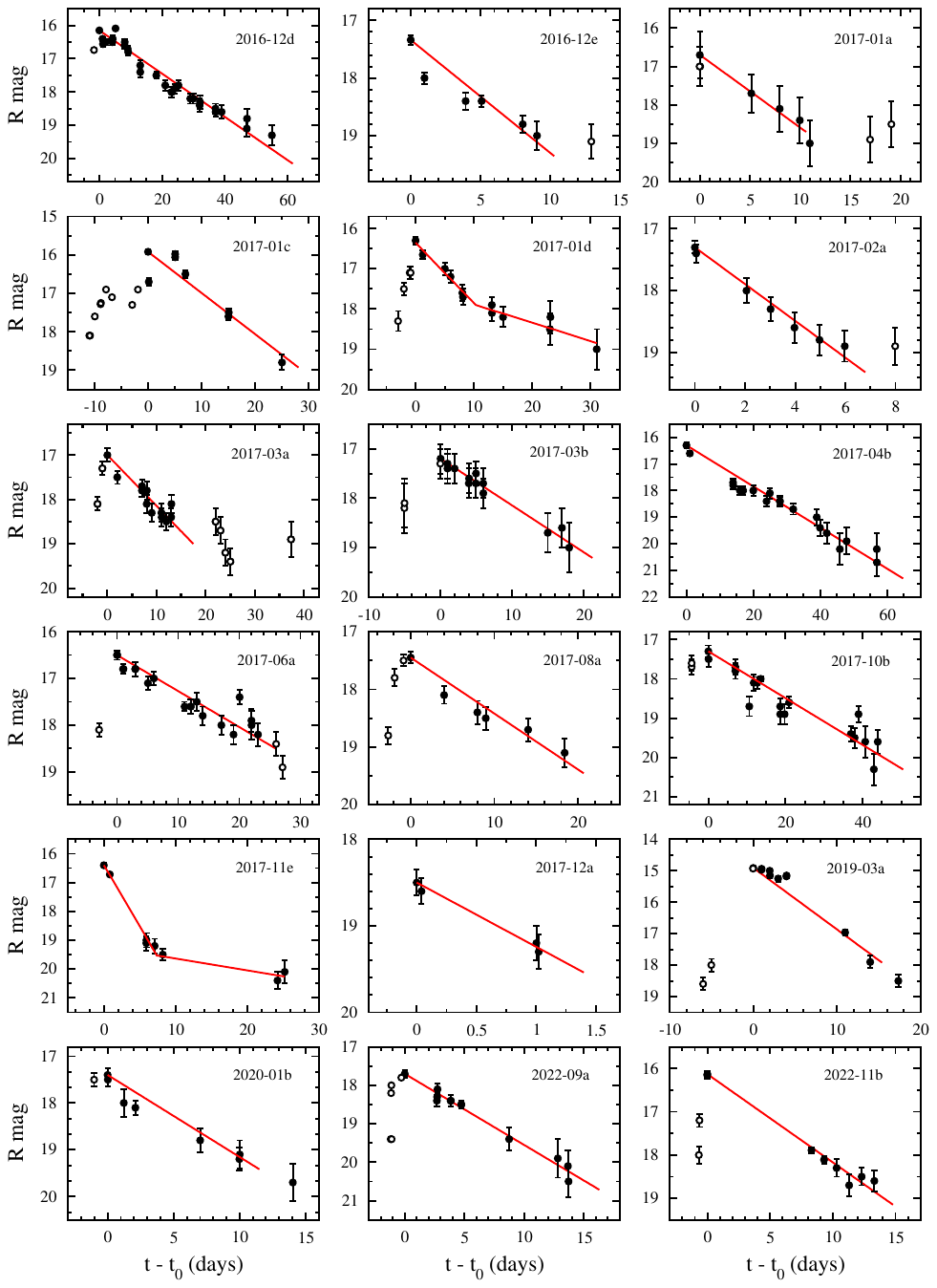}
\caption{Continuation of Figure 1.
}
\label{fig:f1.10}
\end{figure}

\setcounter{figure}{1}

The $R$-band light curves from our complete sample of M31 novae from 2002 to 2022 are presented in Figure~1.
Multicolor light curves of novae covering the years prior to 2010 have been published previously by \citet{2011ApJ...734...12S}
as part of their spectroscopic and photometric survey of novae in M31. However, for completeness, and because we will be applying a consistent methodology to analyze the light curve data across our entire dataset, we have reproduced our earlier $R$-band data in the first
two panels of Figure Set~1.

It has been appreciated for almost a century that novae display a wide range of light curve morphologies \citep[e.g.][]{1936PA.....44...78G,1939PA.....47..410M}. Not only does the
evolution of nova luminosity vary significantly between systems, with some novae fading by up to three magnitudes from peak within a few days \citep[e.g., $t_3\sim2.2$~d for V1674 Her,][]{2021RNAAS...5..160Q} while others can stay within 3 mag of peak luminosity for up to a year, or longer\footnote{Related systems, the so-called ``Symbiotic Novae", containing evolved, pulsationally-unstable (Mira) secondary stars, can remain near maximum light for of order a decade \citep[e.g., see][]{2008JAVSO..36....9S}.}, the declining branch of the eruption itself exhibits a complex structure that varies markedly from one system to another. Perhaps the most extreme examples are provided by the dust forming novae such as DQ Her \citep[][]{2014ASPC..490..261H} where extinction from the newly formed dust results in a deep dip in the declining branch of the optical light curve.

Relatively recently, \citet{2010AJ....140...34S} has proposed dividing nova light curves into a number of distinct morphological types. Many of these classes, which have all been devised from the analysis of Galactic novae, can only be distinguished for novae with light curves that are well sampled and extend to $\grtsim3$ mag below maximum light. In examining our sample of M31 novae -- many of which have relatively fragmentary temporal coverage -- we have found it convenient to divide the light curves into three broad classes: (1) A ``Linear" class that includes light curves showing an approximately linear decline (exponential in flux) from maximum light, (2) A``Break" class composed of light curves displaying a pronounced change in slope on the declining branch of the eruption,
and finally (3) a ``Jitter" class for light curves that exhibit ill-defined maxima and erratic declines, with evidence of multiple peaks, or rebrightenings. Novae
in our Linear and Break classes roughly correspond to the ``S" and ``P" classes, respectively, from \citet{2010AJ....140...34S}. It is possible that some novae in our Linear class could be candidates for the Break class if the light curves were followed for a sufficient length of time. In fact, it may be the case that most if not all novae
could best be fit by a two-component decline if the light curve were followed sufficiently long after eruption. Thus, our distinction of the Linear and Break classes may not reflect a fundamental difference in the evolution of most novae. Objects in the third class are almost exclusively relatively slow novae that can take several months to decline by two magnitudes from peak brightness. These novae broadly fit into the J or ``Jitter" group in the Strope et al. classification scheme.
Essentially all of the known recurrent novae, most of which fade quickly, appear to be consistent with inclusion into the Linear class.

\subsection{Determining Peak magnitudes and fade rates}

Nova light curves can be most simply characterized by three parameters: (1) the rise time to peak brightness, $t_\mathrm{rise}$, (2) the peak magnitude, $m_\mathrm{peak}$, and (3) the fade rate from peak brightness, $f$. The stochastic nature of nova eruptions coupled with the rapid rise of a nova to maximum light (typically of order a day or so) makes it difficult to reliably measure either the rise time or the peak nova brightness. For the latter, often we
must rely on the brightness at discovery to be the best estimate of the magnitude at the peak of the eruption. Clearly, the peak magnitude determined in the manner is strictly speaking a lower limit to the flux (i.e., an upper limit
to the magnitude) at actual maximum light. A good example is the light curve for M31N 2013-10c where the unfiltered discovery magnitude from LOSS was not included in our $R$-band dataset \citep[][]{2024RNAAS...8....5S}.
On the other hand, the rate of decline from peak brightness can be readily computed from a linear fit to the declining branch of the eruption. Since the zero point (intercept) of a general linear fit often underestimates the observed peak brightness (and the fade rate), we have constrained our linear fits to pass through the observed peak magnitude (which usually corresponds to the magnitude at discovery) and fitted for the slope, $f$, of the decline alone.
Assigning peak brightness following this procedure helps to minimize any potential
discrepancy between the observed and actual peak magnitude. Following the traditional practice, in the analysis to follow we will characterize the fade rate by the traditional
$t_2$ time, which is the time in days it takes a nova to fade
2 magnitudes from maximum light ($m_\mathrm{peak+2}$). We note that given the complex structure of most nova light curves, our estimates of
the $t_2$ times for many novae will have true uncertainties that exceed the formal values determined in the fits.

Tables~1--4 give the light curve parameters for our complete sample of novae, where
we have differentiated between the Linear, Break, Jitter, and known recurrent nova classes, respectively. We note that this
distinction was not observed in the earlier \citet{2011ApJ...734...12S} analysis of the pre-2010 data.
Thus, for consistency, we have re-determined the light curve parameters for these novae following the
criteria described above. As a result, the pre-2010 light curve parameters
given in Tables 1--4 can be expected to differ somewhat from those given in the earlier paper.
For the novae in the Break class the light curves are characterized by two fade rates, $f_1$ and $f_2$, determined from fits on either side of the
break point (defined by $m_\mathrm{break}$ and $t_\mathrm{break}$).
A proper determination of the
$t_2$ time for these novae is complicated by any change in slope that occurs before the nova has faded by 2 magnitudes from maximum light (i.e., if $m_\mathrm{break} < m_\mathrm{peak+2}$), and thus is not simply given by $t_2=2/f_1$. Instead, in such cases the time where the light curve
crosses the $m_\mathrm{peak+2}$ point occurs on the second slope, and it is given by $t_2 = t_\mathrm{break} + (m_\mathrm{peak+2} - m_\mathrm{break})/f_2$.
The peak magnitudes have been converted to absolute $R$ magnitudes by adopting a distance modulus and foreground $R$-band extinction for M31 of
$\mu_0 = 24.42\pm0.06$ \citep{2013AJ....146...86T} and $A_R=0.14$ mag \citep{2011ApJ...737..103S}.

Below we explore the relationship between light curve parameters by considering
the Maximum Magnitude, Rate of Decline (MMRD) relation.

\section{The MMRD Relation for M31} \label{sec:MMRD}

\begin{figure*}
\plotone{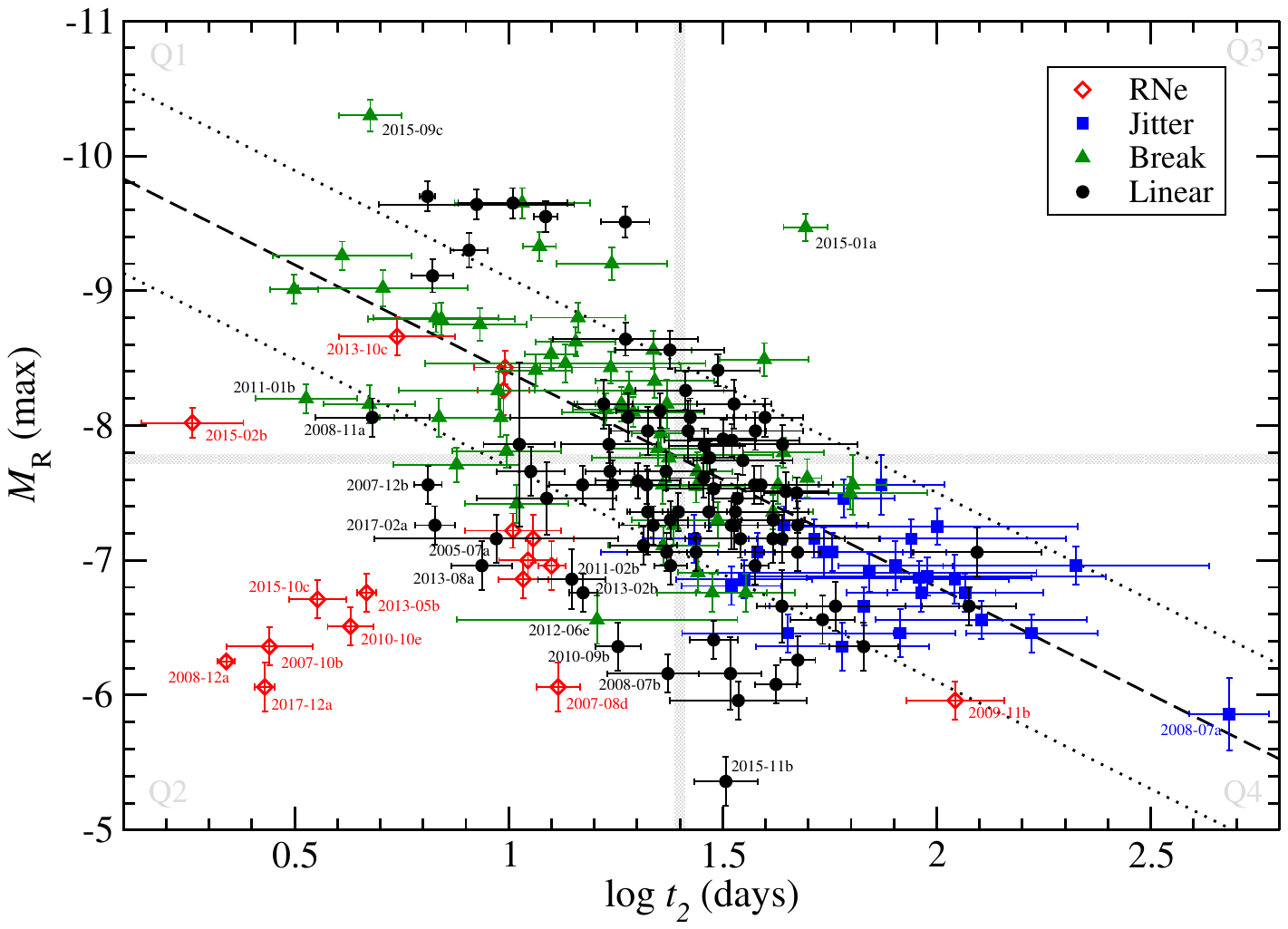}
\caption{The MMRD relation for M31. The data show a weak correlation between the absolute magnitude of a nova at the peak of its eruption
and the time for the nova to fade two magnitudes from maximum light.
Following \citet{2023RNAAS...7..191S}, the MMRD plane is divided into four quadrants as indicated
by the light gray lines. Linear, Break, and Jitter class novae as well as the known recurrent novae
are distinguished as indicated in the Key to the figure. Broadly speaking,
the Break and Jitter class novae are found primarily among the fastest and slowest novae, respectively, with the known recurrent systems being mostly confined to the lower left quadrant (Q2) of the MMRD plane. 
The dashed line represents the best fitting linear MMRD relation (excluding known recurrent novae) with
the dotted lines showing $\pm1\sigma$ departures from this relation.
Novae that significantly depart from the MMRD relation (e.g., known and suspected recurrent novae) are identified.
}
\label{fig:f2}
\end{figure*}

Figure~\ref{fig:f2} shows the MMRD relation for our complete sample of M31 novae. Despite considerable scatter in the relation as expected, it appears that the peak absolute magnitude is nevertheless weakly correlated with log~$t_2$. Novae that reach the highest peak luminosity do generally evolve more rapidly compared with the less luminous systems.
As described by \citet{2023RNAAS...7..191S} and discussed further below, recurrent novae (red points in Figure~\ref{fig:f2}) deviate systematically from the MMRD relation, as expected, being found predominately in the lower left quadrant of the MMRD plane.

The MMRD relation for Galactic novae has traditionally taken a linear form:
\begin{equation}
    M_\lambda(\mathrm{max}) = a + b~\mathrm{log}~t_n,
\end{equation}
where $M_\lambda(\mathrm{max})$ is the absolute magnitude at maximum light as measured through a given
photometric filter denoted by $\lambda$ (typically $V$ or $R$), $t_n$ is the time in days for a nova to decline by $n$ (typically 2 or 3) mag from maximum light, and $a$ and $b$ are fitting parameters representing the intercept and slope of the linear MMRD relation.
More complicated representations parameterized by an {\tt arctan} function have been used in some extragalactic nova studies
\citep[e.g., see][]{1994A&A...287..403D} but its use is not justified here given the large scatter seen in our M31 nova sample.

The broken line shown in Figure~\ref{fig:f2} represents the optimum linear fit to our M31 data with the dotted lines showing $\pm1\sigma$
confidence limits about this relation.
In determining the fit we have opted to omit the known recurrent nova systems (red circles), which are known to depart from the MMRD relation. 
The resulting best-fit solution to our M31 data is given by:
\begin{equation}
    M_R = -9.99(\pm0.22) + 1.59(\pm0.15)~\mathrm{log}~t_2.
\end{equation}

The large scatter in the relation makes it clear that the MMRD relation represents, at best, a weak correlation between the peak absolute magnitude of a nova and its rate of decline from maximum light. For a given value of $\log t_2$ the RMS spread is $\sim0.7$ mag. As a result, the measurement of the peak brightness and fade rate of a single extragalactic nova would not allow a distance measurement to better than $\pm\sim$30\%.

\begin{figure*}
\plotone{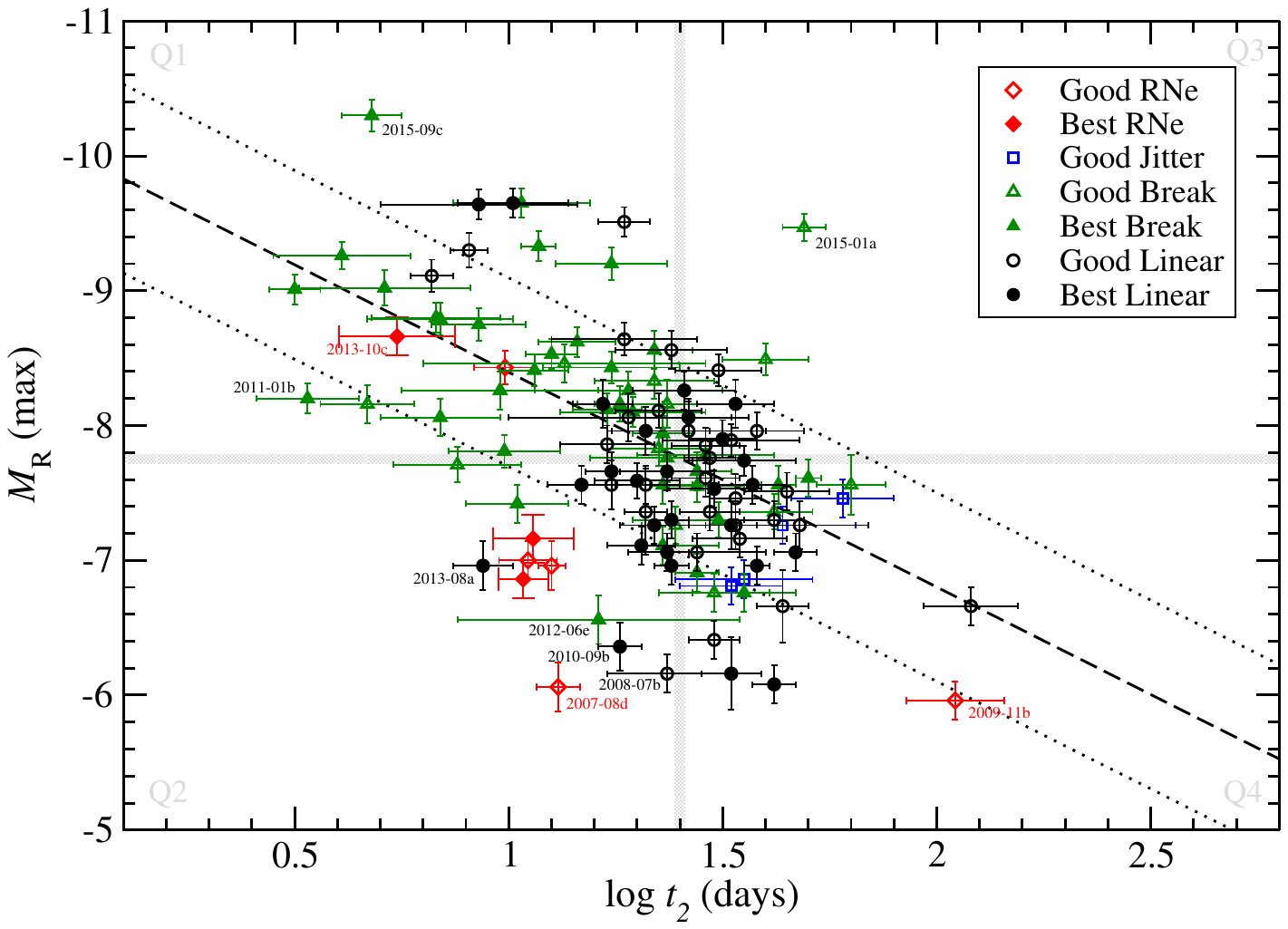}
\caption{The MMRD relation for M31 based on the highest quality data from our Gold and Silver samples. Almost all of the novae in the Jitter class have been omitted as the large amplitude fluctuations made the determination of the fade rate particularly difficult. Restricting the sample to the novae with the most well behaved light curves does not strengthen the MMRD relation. The fits shown (dashed and dotted lines) are reproduced from Figure~\ref{fig:f2}.
}
\label{fig:f3}
\end{figure*}

\subsection{The High Quality MMRD Samples}

We considered the possibility that the MMRD relation might be improved if we restricted our analysis
to novae with the most complete and well behaved light curves where our estimates of the peak magnitudes and fade rates should be the most reliable. From a visual inspection of the light curves for all of our novae we selected 118 high-quality light curves.
We have further distilled this sample to identify 67 novae where we have the highest confidence
in the derived light curve parameters. Following the convention of \citet{2018MNRAS.481.3033S}, these novae having the best determined parameters make up our ``Gold" sample, with the remaining 51 novae
with high quality light curves making up our ``Silver" sample. All remaining novae comprise our ``Bronze" nova sample.

The resulting MMRD relation for the highest quality nova samples is shown in Figure~\ref{fig:f3}.
Despite our expectation that restricting our analysis to these novae would improve the observed
correlation between peak magnitude and fade rate, no improvement is apparent. Indeed, the elimination
of most of the novae in the Jitter class, which have the most erratic light curve declines,
dramatically reduced the number of novae in the lower right quadrant of the MMRD plane. The inclusion
of these slowly fading novae are integral to the character of the MMRD relation, and without them, the MMRD relation becomes much less well defined.

\subsection{The MMRD Relation as a Selection Effect}

\citet{2023RNAAS...7..191S} has recently argued that the observed MMRD relation results from the expectations of nova theory coupled
with observational selection biases.
Following \citet{2023RNAAS...7..191S} it is useful to divide the MMRD plane into four quadrants: (Q1) fast and bright novae in the
upper left, (Q2) fast and faint novae in the lower left, (Q3) slow and bright novae in the upper right, and finally (Q4) slow and faint novae in the lower right quadrant. Where a given nova will fall in the MMRD plane ultimately depends on the
properties of the progenitor binary, principally the white dwarf mass ($M_\mathrm{WD}$) and the rate of accretion onto its surface. These
parameters are important both because they determine amount of mass ($M_\mathrm{ign}$) that must be accreted before a thermonuclear runaway (TNR) is triggered, and because they determine the degeneracy of the accreted envelope (and thus the strength of the TNR)
\citep[e.g.,][]{1982ApJ...253..798N,2005ApJ...628..395T,2014ApJ...793..136K}.

Models show that $M_\mathrm{ign}$ is the smallest
for systems having massive WDs accreting at a high rate, while it is the highest for systems with low mass white dwarfs
accreting a low rate. Thus, the time taken for a given system to achieve a TNR will (the recurrence time) be the shortest ($\lessim10^2$~yr) for
the high mass, high accretion rate systems. These are the recurrent novae. Conversely,
the low mass, low accretion rate systems are expected to have extremely long recurrence times, of order $10^5$~yr \citep[e.g., see][their figure 6]{2014ApJ...793..136K}. In addition, high accretion rates in the recurrent novae can be expected to suppress the degeneracy in the accreted layer resulting in weaker TNRs. Thus, although the recurrent novae will erupt much more frequently than the low accretion rate systems, they will result in
relatively weak, low luminosity, eruptions. Finally, the small accreted mass in the recurrent nova systems will translate into correspondingly lower expelled masses, which in turn will result the ejected gas becoming optically thin soon after the eruption leading to a more rapid photometric evolution. The net result is that recurrent novae will be characterized by``fast and faint" eruptions that lie in the lower left quadrant
of the MMRD plane. Referring to Figure~\ref{fig:f2}, that is precisely what is observed.

What about the progenitors harboring slowly accreting, low mass WDs that must accrete a massive envelope prior to ignition of the TNR?
These systems, which erupt with recurrence timescales thousands of time longer than the known recurrent novae, are
expected to be extremely rare. When the eruption finally does occur it will do so in a highly degenerate layer. The resulting eruption will therefore be luminous and can be expected to produce a massive ejecta that evolves relatively slowly. Thus, these novae will reside in the sparsely populated upper right quadrant (Q3) of the MMRD plane. A lone example of such a nova is M31N 2015-01a. The paucity of novae observed in this quadrant can be understood simply as a result of the extremely long recurrence time expected for progenitors with slowly accreting, low mass white dwarfs. On the other hand, it is less obvious why the lower left quadrant (Q2) should be relatively devoid of novae. After all, novae in this quadrant have the shortest recurrence times and should be relatively common. The resolution of this apparent contradiction is that the discovery of the fast and faint novae in Q2 has been strongly selected against in legacy magnitude-limited surveys that were plagued with poor temporal resolution. These earlier surveys therefore were primarily
sensitive to novae arising in the first and fourth quadrants of the MMRD plane. The bright and rapidly evolving systems in Q1 predominately arise
from progenitors with relatively high mass white dwarfs accreting at a modest rate, while the faint and slowly evolving novae in Q4 are
produced by progenitors having low-mass WDs accreting at relatively high rates. Given the paucity of novae either detected or produced (in Q2
and Q3, respectively), we are left with novae primarily inhabiting the first and fourth quadrants of the MMRD plane. This provides a general framework by which the traditional MMRD relation can be understood. 

\startlongtable
\begin{deluxetable*}{lrccrrcc}
\tablenum{1}
\tablecolumns{8}
\tabletypesize{\scriptsize}
\tablecaption{Linear Nova Light Curve Parameters\label{tab1}}
\tablehead{\colhead{Nova} & & \colhead{JD (max)} &  &  & \colhead{$t_2$} & & \\ \colhead{(M31N)} & \colhead{Sample} & \colhead{(2,450,000+)} & \colhead{$m_R$ (max)} & \colhead{$M_R$ (max)} & \colhead{(d)} & \colhead{log $t_2$} & \colhead{$M_{15}$}}
\startdata
2002-08a & Silver & 2490.523 & $  17.05\pm 0.10$&$ -7.51\pm 0.14$&$  44.4\pm10.3$& $ 1.65\pm 0.10$&$ -6.83\pm 0.21$ \cr
2004-08b & Silver & 3225.482 & $  17.30\pm 0.15$&$ -7.26\pm 0.18$&$  47.4\pm18.0$& $ 1.68\pm 0.16$&$ -6.63\pm 0.30$ \cr
2004-09a & Silver & 3251.518 & $  17.50\pm 0.10$&$ -7.06\pm 0.14$&$  27.4\pm10.1$& $ 1.44\pm 0.16$&$ -5.96\pm 0.43$ \cr
2004-11a & Silver & 3315.390 & $  16.50\pm 0.15$&$ -8.06\pm 0.18$&$  19.0\pm12.0$& $ 1.28\pm 0.28$&$ -6.48\pm 1.02$ \cr
2004-11b & Silver & 3315.390 & $  16.60\pm 0.10$&$ -7.96\pm 0.14$&$  37.7\pm 9.7$& $ 1.58\pm 0.11$&$ -7.16\pm 0.25$ \cr
2005-01a & Silver & 3384.212 & $  15.05\pm 0.05$&$ -9.51\pm 0.11$&$  18.7\pm 2.4$& $ 1.27\pm 0.06$&$ -7.91\pm 0.24$ \cr
2005-07a & Bronze & 2581.419 & $  17.40\pm 0.15$&$ -7.16\pm 0.18$&$   9.4\pm 6.1$& $ 0.97\pm 0.28$&$ -3.95\pm 2.11$ \cr
2006-06a & Gold & 3892.518 & $  17.60\pm 0.10$&$ -6.96\pm 0.14$&$  37.6\pm 2.8$& $ 1.58\pm 0.03$&$ -6.16\pm 0.15$ \cr
2006-09c & Gold & 4000.317 & $  17.00\pm 0.10$&$ -7.56\pm 0.14$&$  14.9\pm 2.7$& $ 1.17\pm 0.08$&$ -5.54\pm 0.39$ \cr
2006-12a & Gold & 4093.174 & $  17.30\pm 0.15$&$ -7.26\pm 0.18$&$  33.1\pm 4.6$& $ 1.52\pm 0.06$&$ -6.35\pm 0.22$ \cr
2007-02b & Gold & 4135.298 & $  16.66\pm 0.10$&$ -7.90\pm 0.14$&$  31.7\pm 5.7$& $ 1.50\pm 0.08$&$ -6.95\pm 0.22$ \cr
2007-07e & Silver & 4327.469 & $  17.90\pm 0.25$&$ -6.66\pm 0.27$&$  43.5\pm 5.8$& $ 1.64\pm 0.06$&$ -5.97\pm 0.28$ \cr
2007-11b & Bronze & 4415.446 & $  18.30\pm 0.15$&$ -6.26\pm 0.18$&$  47.4\pm 4.4$& $ 1.68\pm 0.04$&$ -5.63\pm 0.19$ \cr
2007-12b & Bronze & 4445.234 & $  17.00\pm 0.10$&$ -7.56\pm 0.14$&$   6.5\pm 0.5$& $ 0.81\pm 0.03$&$ -2.93\pm 0.37$ \cr
2008-05c & Silver & 4619.529 & $  17.20\pm 0.10$&$ -7.36\pm 0.14$&$  29.4\pm 5.1$& $ 1.47\pm 0.07$&$ -6.34\pm 0.23$ \cr
2008-06b & Silver & 4644.498 & $  16.00\pm 0.10$&$ -8.56\pm 0.14$&$  23.8\pm 6.9$& $ 1.38\pm 0.13$&$ -7.30\pm 0.39$ \cr
2008-07b & Silver & 4675.581 & $  18.40\pm 0.10$&$ -6.16\pm 0.14$&$  23.5\pm 7.7$& $ 1.37\pm 0.14$&$ -4.89\pm 0.44$ \cr
2008-08a & Silver & 4692.616 & $  16.70\pm 0.10$&$ -7.86\pm 0.14$&$  17.1\pm 4.4$& $ 1.23\pm 0.11$&$ -6.11\pm 0.47$ \cr
2008-10b & Bronze & 4800.301 & $  18.00\pm 0.15$&$ -6.56\pm 0.18$&$  54.1\pm 9.3$& $ 1.73\pm 0.07$&$ -6.01\pm 0.20$ \cr
2008-11a & Bronze & 4775.218 & $  16.50\pm 0.10$&$ -8.06\pm 0.14$&$   4.8\pm 1.5$& $ 0.68\pm 0.13$&$ -1.81\pm 1.93$ \cr
2009-08d & Bronze & 5055.923 & $  17.20\pm 0.10$&$ -7.36\pm 0.14$&$  33.9\pm13.3$& $ 1.53\pm 0.17$&$ -6.47\pm 0.38$ \cr
2009-08e & Silver & 5073.441 & $  17.90\pm 0.10$&$ -6.66\pm 0.14$&$ 118.9\pm30.3$& $ 2.08\pm 0.11$&$ -6.41\pm 0.16$ \cr
2009-09a & Bronze & 5080.413 & $  17.50\pm 0.15$&$ -7.06\pm 0.18$&$ 124.2\pm43.1$& $ 2.09\pm 0.15$&$ -6.82\pm 0.20$ \cr
2009-10c & Gold & 5124.491 & $  16.40\pm 0.15$&$ -8.16\pm 0.18$&$  33.6\pm 6.7$& $ 1.53\pm 0.09$&$ -7.27\pm 0.25$ \cr
2009-11a & Gold & 5140.463 & $  17.60\pm 0.10$&$ -6.96\pm 0.14$&$  23.9\pm 2.1$& $ 1.38\pm 0.04$&$ -5.70\pm 0.18$ \cr
2009-11c & Gold & 5148.536 & $  17.03\pm 0.10$&$ -7.53\pm 0.14$&$  30.1\pm 7.7$& $ 1.48\pm 0.11$&$ -6.53\pm 0.29$ \cr
2009-11d & Gold & 5157.299 & $  17.26\pm 0.10$&$ -7.30\pm 0.14$&$  23.9\pm 2.2$& $ 1.38\pm 0.04$&$ -6.04\pm 0.18$ \cr
2009-11e & Bronze & 5158.542 & $  17.00\pm 0.10$&$ -7.56\pm 0.14$&$  38.8\pm15.1$& $ 1.59\pm 0.17$&$ -6.79\pm 0.33$ \cr
2010-05a  & Gold & 5353.979 & $ 17.00\pm 0.10$&$ -7.56\pm 0.14$&$  37.5\pm6.8$& $  1.57\pm 0.08$&$ -6.76\pm 0.20$ \cr
2010-09b  & Gold & 5476.432 & $ 18.20\pm0.15$&$ -6.36\pm0.18$&$ 18.0\pm2.2$&  $1.26\pm0.05$&$ -4.69\pm0.27$ \cr
2010-10c  & Gold & 5490.515 & $ 18.15\pm0.10$&$ -6.41\pm0.14$&$ 30.1\pm3.9$&  $1.48\pm0.06$&$ -5.41\pm0.19$ \cr
2011-01a  & Gold & 5573.463 & $ 14.92\pm0.05$&$ -9.64\pm0.11$&$  8.4\pm4.4$&  $0.93\pm0.23$&$ -6.08\pm1.87$ \cr
2011-02a  & Bronze & 5613.236 & $ 17.20\pm0.10$&$ -7.36\pm0.14$&$ 24.9\pm6.5$&  $1.40\pm0.11$&$ -6.16\pm0.34$ \cr
2011-02b  & Bronze & 5616.284 & $ 17.70\pm0.20$&$ -6.86\pm0.22$&$ 14.0\pm2.6$&  $1.15\pm0.08$&$ -4.72\pm0.45$ \cr
2011-02c  & Bronze & 5616.284 & $ 16.90\pm0.15$&$ -7.66\pm0.18$&$ 11.3\pm2.1$&  $1.05\pm0.08$&$ -4.99\pm0.52$ \cr
2011-02d  & Gold & 5621.241 & $ 16.90\pm0.10$&$ -7.66\pm0.14$&$ 23.3\pm5.8$&  $1.37\pm0.11$&$ -6.37\pm0.35$ \cr
2011-06b  & Gold & 5726.634 & $ 18.48\pm0.10$&$ -6.08\pm0.14$&$ 42.1\pm4.7$&  $1.62\pm0.05$&$ -5.37\pm0.16$ \cr
2011-06d  & Gold & 5748.533 & $ 16.90\pm0.10$&$ -7.66\pm0.14$&$ 17.2\pm2.2$&  $1.24\pm0.06$&$ -5.92\pm0.27$ \cr
2011-07a  & Silver & 5765.584 & $ 16.45\pm0.08$&$ -8.11\pm0.13$&$ 22.6\pm5.4$&  $1.35\pm0.10$&$ -6.78\pm0.34$ \cr
2011-07b  & Gold & 5776.596 & $ 17.50\pm0.10$&$ -7.06\pm0.14$&$ 23.4\pm2.6$&  $1.37\pm0.05$&$ -5.78\pm0.20$ \cr
2011-08b  & Gold & 5778.462 & $ 16.82\pm0.04$&$ -7.74\pm0.11$&$ 35.2\pm9.4$&  $1.55\pm0.12$&$ -6.89\pm0.25$ \cr
2011-12b  & Bronze & 5931.290 & $ 18.20\pm0.15$&$ -6.36\pm0.18$&$ 67.5\pm12.5$& $1.83\pm0.08$&$ -5.92\pm0.20$ \cr
2012-01a  & Silver & 5937.217 & $ 17.40\pm0.10$&$ -7.16\pm0.14$&$ 34.8\pm8.7$&  $1.54\pm0.11$&$ -6.30\pm0.26$ \cr
2012-02c  & Silver & 5978.234 & $ 16.95\pm0.10$&$ -7.61\pm0.14$&$ 28.6\pm5.3$&  $1.46\pm0.08$&$ -6.56\pm0.24$ \cr
2012-03b  & Bronze & 6005.273 & $ 17.40\pm0.15$&$ -7.16\pm0.18$&$ 43.5\pm22.7$& $1.64\pm0.23$&$ -6.47\pm0.40$ \cr
2012-03c  & Bronze & 6010.269 & $ 16.50\pm0.10$&$ -8.06\pm0.14$&$ 39.7\pm8.1$&  $1.60\pm0.09$&$ -7.30\pm0.21$ \cr
2012-05c  & Gold & 6053.558 & $ 17.30\pm0.10$&$ -7.26\pm0.14$&$ 21.7\pm4.0$&  $1.34\pm0.08$&$ -5.88\pm0.29$ \cr
2012-05d  & Gold & 6053.567 & $ 18.40\pm0.25$&$ -6.16\pm0.27$&$ 33.0\pm5.6$&  $1.52\pm0.07$&$ -5.25\pm0.31$ \cr
2012-06a  & Silver & 6098.538 & $ 16.67\pm0.07$&$ -7.89\pm0.12$&$ 33.2\pm12.1$& $1.52\pm0.16$&$ -6.99\pm0.35$ \cr
2012-06f  & Bronze & 6101.535 & $ 18.60\pm0.10$&$ -5.96\pm0.14$&$ 34.4\pm12.7$& $1.54\pm0.16$&$ -5.09\pm0.35$ \cr
2012-06g  & Gold & 6105.552 & $ 16.97\pm0.08$&$ -7.59\pm0.13$&$ 20.0\pm4.6$&  $1.30\pm0.10$&$ -6.09\pm0.36$ \cr
2012-07a  & Bronze & 6127.462 & $ 17.90\pm0.15$&$ -6.66\pm0.18$&$ 58.0\pm21.9$& $1.76\pm0.16$&$ -6.14\pm0.27$ \cr
2012-07b  & Silver & 6151.580 & $ 16.80\pm0.07$&$ -7.76\pm0.12$&$ 29.5\pm4.0$&  $1.47\pm0.06$&$ -6.74\pm0.18$ \cr
2013-02b  & Bronze & 6354.246 & $ 17.80\pm0.10$&$ -6.76\pm0.14$&$ 14.9\pm1.1$&  $1.17\pm0.03$&$ -4.75\pm0.21$ \cr
2013-03b  & Bronze & 6376.272 & $ 16.70\pm0.10$&$ -7.86\pm0.14$&$ 43.6\pm17.6$& $1.64\pm0.17$&$ -7.17\pm0.31$ \cr
2013-06a  & Bronze & 6458.554 & $ 17.10\pm0.25$&$ -7.46\pm0.27$&$ 12.3\pm4.6$&  $1.09\pm0.16$&$ -5.01\pm0.96$ \cr
2013-08a  & Gold & 6513.479 & $ 17.60\pm0.15$&$ -6.96\pm0.18$&$  8.7\pm1.4$&  $0.94\pm0.07$&$ -3.49\pm0.60$ \cr
2013-10g  & Bronze & 6592.227 & $ 17.06\pm0.06$&$ -7.50\pm0.12$&$ 47.1\pm8.2$&  $1.67\pm0.08$&$ -6.86\pm0.16$ \cr
2014-01c  & Silver & 6683.314 & $ 17.10\pm0.15$&$ -7.46\pm0.18$&$ 34.2\pm4.6$&  $1.53\pm0.06$&$ -6.58\pm0.22$ \cr
2014-06a  & Gold & 6837.518 & $ 16.60\pm0.15$&$ -7.96\pm0.18$&$ 21.1\pm4.4$&  $1.32\pm0.09$&$ -6.54\pm0.35$ \cr
2014-07a  & Gold & 6851.479 & $ 17.50\pm0.10$&$ -7.06\pm0.14$&$ 47.3\pm5.8$&  $1.67\pm0.05$&$ -6.43\pm0.16$ \cr
2014-09a  & Gold & 6910.327 & $ 16.40\pm0.15$&$ -8.16\pm0.18$&$ 16.7\pm2.6$&  $1.22\pm0.07$&$ -6.36\pm0.34$ \cr
2015-05b  & Silver & 7173.530 & $ 17.00\pm0.10$&$ -7.56\pm0.14$&$ 21.0\pm3.7$&  $1.32\pm0.08$&$ -6.13\pm0.29$ \cr
2015-06a  & Bronze & 7179.524 & $ 14.86\pm0.05$&$ -9.70\pm0.11$&$  6.5\pm0.3$&  $0.81\pm0.02$&$ -5.06\pm0.22$ \cr
2015-07d  & Silver & 7231.527 & $ 15.45\pm0.07$&$ -9.11\pm0.12$&$  6.6\pm0.7$&  $0.82\pm0.05$&$ -4.59\pm0.52$ \cr
2015-08a  & Silver & 7257.615 & $ 17.26\pm0.10$&$ -7.30\pm0.14$&$ 41.5\pm7.1$&  $1.62\pm0.07$&$ -6.58\pm0.19$ \cr
2015-09a  & Bronze & 7280.349 & $ 15.01\pm0.05$&$ -9.55\pm0.11$&$ 12.2\pm0.8$&  $1.09\pm0.03$&$ -7.09\pm0.19$ \cr
2015-10a  & Silver & 7300.262 & $ 15.26\pm0.08$&$ -9.30\pm0.13$&$   8.1\pm0.8$& $0.91\pm0.04$&$  -5.59\pm0.39$ \cr
2015-11b  & Bronze & 7340.508 & $ 19.20\pm0.15$&$ -5.36\pm0.18$&$ 32.2\pm5.5$&  $1.51\pm0.07$&$ -4.43\pm0.24$ \cr
2016-03b  & Bronze & 7456.318 & $ 17.26\pm0.10$&$ -7.30\pm0.14$&$ 41.5\pm7.1$&  $1.62\pm0.07$&$ -6.58\pm0.19$ \cr
2016-10b  & Silver & 7681.329 & $ 16.71\pm0.08$&$ -7.85\pm0.13$&$ 28.6\pm5.5$&  $1.46\pm0.08$&$ -6.80\pm0.24$ \cr
2016-10c  & Silver & 7678.291 & $ 16.60\pm0.20$&$ -7.96\pm0.22$&$ 26.3\pm10.9$& $1.42\pm0.18$&$ -6.82\pm0.52$ \cr
2016-12b  & Bronze & 7753.173 & $ 17.40\pm0.15$&$ -7.16\pm0.18$&$ 41.4\pm16.0$& $1.62\pm0.17$&$ -6.44\pm0.33$ \cr
2016-12d  & Silver & 7751.210 & $ 16.15\pm0.06$&$ -8.41\pm0.12$&$ 30.8\pm7.0$&  $1.49\pm0.10$&$ -7.44\pm0.25$ \cr
2017-01a  & Bronze & 7764.234 & $ 16.70\pm0.60$&$ -7.86\pm0.61$&$ 10.6\pm2.0$&  $1.02\pm0.08$&$ -5.03\pm0.82$ \cr
2017-01c  & Silver & 7783.204 & $ 15.92\pm0.07$&$ -8.64\pm0.12$&$ 18.7\pm7.3$&  $1.27\pm0.17$&$ -7.04\pm0.63$ \cr
2017-02a  & Bronze & 7806.266 & $ 17.30\pm0.10$&$ -7.26\pm0.14$&$  6.7\pm0.7$&  $0.83\pm0.05$&$ -2.80\pm0.51$ \cr
2017-03a  & Silver & 7816.270 & $ 17.00\pm0.15$&$ -7.56\pm0.18$&$ 17.5\pm3.3$&  $1.24\pm0.08$&$ -5.84\pm0.37$ \cr
2017-03b  & Silver & 7823.320 & $ 17.20\pm0.30$&$ -7.36\pm0.32$&$ 21.1\pm2.3$&  $1.32\pm0.05$&$ -5.94\pm0.35$ \cr
2017-04b  & Gold & 7867.608 & $ 16.30\pm0.10$&$ -8.26\pm0.14$&$ 25.9\pm7.0$&  $1.41\pm0.12$&$ -7.10\pm0.34$ \cr
2017-06a  & Gold & 7912.494 & $ 16.50\pm0.10$&$ -8.06\pm0.14$&$ 26.6\pm6.1$&  $1.42\pm0.10$&$ -6.93\pm0.30$ \cr
2017-08a  & Gold & 8043.231 & $ 17.45\pm0.10$&$ -7.11\pm0.14$&$ 20.6\pm3.7$&  $1.31\pm0.08$&$ -5.66\pm0.30$ \cr
2017-10b  & Silver & 8061.693 & $ 17.30\pm0.15$&$ -7.26\pm0.18$&$ 33.6\pm11.6$& $1.53\pm0.15$&$ -6.37\pm0.36$ \cr
2019-03a  & Gold & 8561.301 & $ 14.91\pm0.05$&$ -9.65\pm0.11$&$ 10.2\pm3.0$&  $1.01\pm0.13$&$ -6.72\pm0.87$ \cr
\enddata
\end{deluxetable*}

\startlongtable
\begin{deluxetable*}{lrccrrrrccc}
\tablenum{2}
\tablecolumns{11}
\tabletypesize{\tiny}
\tablecaption{Break Nova Light Curve Parameters\label{tab2}}
\tablehead{\colhead{Nova} & & & & \colhead{} & \colhead{$t_2$} & & \colhead{$t_\mathrm{br}$} & & \colhead{$f_\mathrm{2nd}$} & \\ \colhead{(M31N)} & \colhead{Sample} & \colhead{JD (max)} & \colhead{$m_R$ (max)} & \colhead{$M_R$ (max)} & \colhead{(d)} & \colhead{log $t_2$} & \colhead{(d)} & \colhead{$m_{R,\mathrm{br}}$} & \colhead{(mag~d$^{-1}$)} & \colhead{$M_{15}$}}
\startdata
2006-11a & Gold & 4078.343&$16.00\pm0.10$&$ -8.56\pm0.14$&$21.8\pm4.5$&$1.34\pm0.09$&$30.1\pm4.4$&$18.92\pm0.24$&$ 0.012\pm0.012$&$-7.18\pm0.30$ \cr
2009-10b & Silver & 5124.460&$15.76\pm0.04$&$ -8.80\pm0.11$&$ 6.8\pm2.3$&$0.83\pm0.15$&$ 7.0\pm1.9$&$17.84\pm0.28$&$ 0.028\pm0.009$&$-6.49\pm0.31$ \cr
2010-01a & Gold & 5209.610&$15.94\pm0.04$&$ -8.62\pm0.11$&$14.3\pm 3.1$&$ 1.16\pm 0.09$&$17.7\pm 2.5$&$ 18.75\pm0.29$&$ 0.019\pm 0.032$&$ -6.53\pm0.46$ \cr
2010-01b & Gold & 5219.226 &$ 16.80\pm  0.15$&$ -7.76\pm  0.18$&$  23.7\pm   9.7$&$  1.37\pm  0.18$&$   4.1\pm   2.9$&$ 17.69\pm  0.30$&$ 0.057\pm 0.021$&$ -6.25\pm  0.42$ \cr
2010-01c & Bronze & 5226.321&$ 16.85\pm0.08$&$ -7.71\pm0.13$&$ 7.6\pm 2.6$&$ 0.88\pm 0.15$&$ 6.6\pm 1.7$&$ 18.55\pm0.33$&$ 0.009\pm 0.011$&$ -5.93\pm0.36$ \cr
2010-02a& Gold & 5255.249&$ 15.54\pm  0.09$&$ -9.02\pm  0.13$&$   5.1\pm   2.3$&$  0.71\pm  0.20$&$   2.4\pm   0.9$&$ 16.96\pm  0.35$&$ 0.213\pm 0.102$&$ -4.90\pm  1.35$ \cr
2010-03a & Silver & 5260.273&$ 17.00\pm0.20$&$ -7.56\pm0.22$&$63.7\pm11.6$&$ 1.80\pm 0.08$&$79.5\pm11.8$&$ 19.74\pm0.29$&$ 0.015\pm 0.010$&$ -7.09\pm0.13$ \cr
2010-06a & Gold & 5377.522&$ 16.95\pm0.10$&$ -7.61\pm0.14$&$49.8\pm 3.9$&$ 1.70\pm 0.03$&$73.3\pm 3.6$&$ 20.31\pm0.12$&$ 0.013\pm 0.007$&$ -7.01\pm0.11$ \cr
2010-06c & Gold & 5376.522&$ 17.45\pm  0.10$&$ -7.11\pm  0.14$&$  23.0\pm   6.8$&$  1.36\pm  0.13$&$  11.1\pm   2.2$&$ 18.78\pm  0.21$&$ 0.056\pm 0.024$&$ -5.56\pm  0.28$ \cr
2010-07a & Gold & 5406.544&$ 16.13\pm  0.06$&$ -8.43\pm  0.12$&$  17.3\pm   6.6$&$  1.24\pm  0.16$&$  15.8\pm   3.2$&$ 18.09\pm  0.15$&$ 0.028\pm 0.004$&$ -6.57\pm  0.62$ \cr
2010-10a & Gold & 5477.250&$ 17.14\pm0.10$&$ -7.42\pm0.14$&$10.4\pm 2.9$&$ 1.02\pm 0.12$&$10.2\pm 2.1$&$ 19.15\pm0.24$&$ 0.029\pm 0.010$&$ -5.27\pm0.27$ \cr
2010-10d & Gold & 5501.261&$ 17.00\pm0.10$&$ -7.56\pm0.14$&$22.9\pm 3.9$&$ 1.36\pm 0.07$&$34.1\pm 1.2$&$ 20.37\pm0.28$&$ 0.010\pm 0.017$&$ -6.25\pm0.24$ \cr
2010-12c & Gold & 5547.229&$ 16.50\pm0.10$&$ -8.06\pm0.14$&$ 6.9\pm 2.2$&$ 0.84\pm 0.14$&$ 6.9\pm 1.8$&$ 18.56\pm0.32$&$ 0.098\pm 0.045$&$ -5.21\pm0.53$ \cr
2011-01b & Gold & 5578.225&$ 16.36\pm0.04$&$ -8.20\pm0.11$&$ 3.4\pm 0.9$&$ 0.53\pm 0.12$&$ 3.7\pm 0.5$&$ 18.82\pm0.30$&$ 0.265\pm 0.123$&$ -2.75\pm1.43$ \cr
2011-06a & Gold & 5727.539&$ 17.80\pm0.10$&$ -6.76\pm0.14$&$35.8\pm 9.5$&$ 1.55\pm 0.12$&$37.6\pm 6.1$&$ 20.02\pm0.27$&$ 0.019\pm 0.010$&$ -5.92\pm0.24$ \cr
2011-08a & Gold & 5778.525&$ 17.65\pm  0.10$&$ -6.91\pm  0.14$&$  27.7\pm   3.0$&$  1.44\pm  0.05$&$  25.8\pm   2.4$&$ 19.25\pm  0.32$&$ 0.220\pm 0.107$&$ -6.02\pm  0.43$ \cr
2011-09a & Gold & 5817.438&$ 16.46\pm  0.05$&$ -8.10\pm  0.11$&$  19.5\pm   7.4$&$  1.29\pm  0.17$&$  15.0\pm   2.7$&$ 18.31\pm  0.22$&$ 0.034\pm 0.013$&$ -6.25\pm  0.26$ \cr
2011-09b & Gold & 5819.271&$ 16.75\pm  0.07$&$ -7.81\pm  0.12$&$   9.9\pm   2.9$&$  0.99\pm  0.13$&$   3.9\pm   1.2$&$ 18.10\pm  0.24$&$ 0.109\pm 0.024$&$ -5.25\pm  0.40$ \cr
2011-10a & Bronze & 5844.489&$ 16.76\pm0.05$&$ -7.80\pm0.11$&$43.9\pm 9.4$&$ 1.64\pm 0.09$&$42.7\pm 5.4$&$ 19.18\pm0.22$&$ 0.016\pm 0.017$&$ -7.12\pm0.18$ \cr
2011-10d & Gold & 5856.607&$ 16.78\pm  0.06$&$ -7.78\pm  0.12$&$  29.0\pm  10.4$&$  1.46\pm  0.16$&$  17.9\pm   4.5$&$ 18.42\pm  0.24$&$ 0.032\pm 0.016$&$ -6.35\pm  0.51$ \cr
2011-11c & Gold & 5888.185&$ 16.62\pm0.08$&$ -7.94\pm0.13$&$22.7\pm 3.6$&$ 1.36\pm 0.07$&$33.1\pm 0.3$&$ 19.78\pm0.22$&$ 0.004\pm 0.014$&$ -6.62\pm0.23$ \cr
2011-11e & Silver & 5901.182&$ 16.07\pm  0.07$&$ -8.49\pm  0.12$&$  39.5\pm   9.5$&$  1.60\pm  0.10$&$   7.3\pm   3.3$&$ 16.95\pm  0.20$&$ 0.035\pm 0.007$&$ -7.34\pm  0.26$ \cr
2011-12a & Gold & 5905.415&$ 17.26\pm  0.09$&$ -7.30\pm  0.13$&$  30.8\pm  14.3$&$  1.49\pm  0.20$&$   1.9\pm   0.8$&$ 18.60\pm  0.24$&$ 0.023\pm 0.007$&$ -5.66\pm  0.27$ \cr
2012-02b & Gold & 5977.262&$ 15.30\pm0.03$&$ -9.26\pm0.10$&$ 4.1\pm 1.5$&$ 0.61\pm 0.16$&$ 4.6\pm 1.2$&$ 17.42\pm0.38$&$ 0.087\pm 0.060$&$ -6.24\pm0.74$ \cr
2012-05a & Gold & 6058.570&$ 15.36\pm  0.07$&$ -9.20\pm  0.12$&$  17.4\pm   5.2$&$  1.24\pm  0.13$&$  10.7\pm   2.7$&$ 17.04\pm  0.20$&$ 0.048\pm 0.006$&$ -7.32\pm  0.26$ \cr
2012-06e & Gold & 6095.554&$ 18.00\pm  0.15$&$ -6.56\pm  0.18$&$  16.1\pm  12.1$&$  1.21\pm  0.33$&$   4.2\pm   1.7$&$ 19.32\pm  0.32$&$ 0.057\pm 0.050$&$ -4.62\pm  0.64$ \cr
2012-07c & Gold & 6133.470&$ 17.01\pm  0.07$&$ -7.55\pm  0.12$&$  27.4\pm   7.2$&$  1.44\pm  0.11$&$  15.1\pm   2.2$&$ 18.45\pm  0.20$&$ 0.046\pm 0.019$&$ -6.13\pm  0.51$ \cr
2012-09b & Gold & 6188.281&$ 15.55\pm0.04$&$ -9.01\pm0.11$&$ 3.2\pm 0.4$&$ 0.50\pm 0.06$&$ 3.7\pm 0.6$&$ 18.03\pm0.23$&$ 0.074\pm 0.018$&$ -5.69\pm0.33$ \cr
2013-04a & Gold & 6390.640&$ 16.30\pm  0.10$&$ -8.26\pm  0.14$&$   9.4\pm   5.1$&$  0.98\pm  0.23$&$   6.7\pm   2.4$&$ 17.98\pm  0.45$&$ 0.114\pm 0.079$&$ -5.63\pm  0.85$ \cr
2013-06b & Silver & 6472.525&$ 16.23\pm0.08$&$ -8.33\pm0.13$&$21.9\pm 7.0$&$ 1.34\pm 0.14$&$18.7\pm 3.1$&$ 18.51\pm0.15$&$ 0.020\pm 0.005$&$ -6.96\pm0.45$ \cr
2013-08b & Silver & 6518.512&$ 16.40\pm  0.15$&$ -8.16\pm  0.18$&$  23.4\pm   9.0$&$  1.37\pm  0.17$&$   8.6\pm   2.4$&$ 18.02\pm  0.16$&$ 0.026\pm 0.004$&$ -6.38\pm  0.20$ \cr
2013-09a & Silver & 6553.377&$ 17.80\pm  0.10$&$ -6.76\pm  0.14$&$  29.9\pm   8.8$&$  1.48\pm  0.13$&$  11.2\pm   2.4$&$ 19.18\pm  0.17$&$ 0.033\pm 0.011$&$ -5.26\pm  0.22$ \cr
2013-09c & Gold & 6553.377&$ 16.44\pm0.06$&$ -8.12\pm0.12$&$16.8\pm 3.0$&$ 1.23\pm 0.08$&$23.2\pm 1.9$&$ 19.48\pm0.18$&$ 0.023\pm 0.021$&$ -6.34\pm0.33$ \cr
2013-09d & Bronze & 6562.256 &$ 17.06\pm  0.07$&$ -7.50\pm  0.12$&$  62.7\pm  26.0$&$  1.80\pm  0.18$&$   1.9\pm   0.7$&$ 18.41\pm  0.16$&$ 0.011\pm 0.003$&$ -6.01\pm  0.20$ \cr
2013-10a & Gold & 6570.239&$ 16.40\pm  0.08$&$ -8.16\pm  0.13$&$  18.3\pm   4.3$&$  1.26\pm  0.10$&$   2.3\pm   1.4$&$ 17.00\pm  0.22$&$ 0.087\pm 0.017$&$ -6.45\pm  0.35$ \cr
2013-10e& Bronze & 6582.516&$ 16.50\pm  0.10$&$ -8.06\pm  0.14$&$   9.6\pm   5.5$&$  0.98\pm  0.25$&$   5.4\pm   1.3$&$ 18.23\pm  0.28$&$ 0.066\pm 0.046$&$ -5.70\pm  0.54$ \cr
2013-10h & Bronze & 6604.192&$ 15.76\pm0.05$&$ -8.80\pm0.11$&$14.5\pm 3.7$&$ 1.16\pm 0.11$&$17.9\pm 2.7$&$ 18.82\pm0.22$&$ 0.043\pm 0.013$&$ -6.73\pm0.54$ \cr
2013-12b & Gold & 6645.244&$ 15.77\pm  0.07$&$ -8.79\pm  0.12$&$   7.0\pm   2.8$&$  0.84\pm  0.17$&$   3.4\pm   1.1$&$ 17.39\pm  0.25$&$ 0.108\pm 0.023$&$ -5.92\pm  0.40$ \cr
2014-01a & Gold & 6661.255&$ 16.15\pm0.06$&$ -8.41\pm0.12$&$11.6\pm 2.2$&$ 1.06\pm 0.08$&$14.6\pm 2.1$&$ 18.77\pm0.27$&$ 0.040\pm 0.015$&$ -5.77\pm0.30$ \cr
2014-06b & Gold & 6842.438&$ 16.90\pm0.10$&$ -7.66\pm0.14$&$27.6\pm 5.2$&$ 1.44\pm 0.08$&$30.5\pm 2.9$&$ 19.07\pm0.13$&$ 0.013\pm 0.003$&$ -6.57\pm0.23$ \cr
2014-12a & Gold & 6999.541&$ 14.91\pm  0.05$&$ -9.65\pm  0.11$&$  10.8\pm   3.9$&$  1.03\pm  0.16$&$   7.5\pm   1.5$&$ 16.69\pm  0.23$&$ 0.070\pm 0.029$&$ -7.35\pm  0.35$ \cr
2015-01a & Silver & 7044.189&$ 15.09\pm  0.01$&$ -9.47\pm  0.10$&$  49.4\pm   5.9$&$  1.69\pm  0.05$&$  40.6\pm   4.1$&$ 16.14\pm  0.30$&$ 0.108\pm 0.039$&$ -9.03\pm  0.28$ \cr
2015-09c & Gold & 7295.283&$ 14.26\pm0.06$&$-10.30\pm0.12$&$ 4.7\pm 0.8$&$ 0.68\pm 0.07$&$ 6.1\pm 0.7$&$ 17.16\pm0.15$&$ 0.032\pm 0.003$&$ -7.12\pm0.18$ \cr
2015-09d & Silver & 7295.498&$ 16.10\pm  0.10$&$ -8.46\pm  0.14$&$  13.6\pm  10.2$&$  1.13\pm  0.33$&$   4.4\pm   1.2$&$ 17.33\pm  0.27$&$ 0.084\pm 0.088$&$ -6.34\pm  0.98$ \cr
2016-05a & Silver & 7519.566&$ 17.30\pm0.10$&$ -7.26\pm0.14$&$24.3\pm 4.5$&$ 1.39\pm 0.08$&$24.6\pm 3.1$&$ 19.57\pm0.16$&$ 0.013\pm 0.005$&$ -6.03\pm0.25$ \cr
2016-08b & Gold & 7612.353&$ 17.00\pm0.10$&$ -7.56\pm0.14$&$42.6\pm 4.3$&$ 1.63\pm 0.04$&$50.4\pm 2.7$&$ 19.38\pm0.11$&$ 0.025\pm 0.012$&$ -6.86\pm0.12$ \cr
2016-08d & Gold & 7622.459&$ 15.23\pm  0.04$&$ -9.33\pm  0.11$&$  11.8\pm   1.1$&$  1.07\pm  0.04$&$   7.0\pm   0.8$&$ 16.04\pm  0.16$&$ 0.246\pm 0.010$&$ -6.54\pm  0.28$ \cr
2016-11a & Gold & 7698.259&$ 15.81\pm0.07$&$ -8.75\pm0.12$&$ 8.6\pm 2.2$&$ 0.93\pm 0.11$&$10.6\pm 1.4$&$ 18.39\pm0.22$&$ 0.049\pm 0.020$&$ -5.95\pm0.27$ \cr
2016-11b & Gold & 7721.326&$ 16.03\pm0.05$&$ -8.53\pm0.11$&$12.6\pm 1.8$&$ 1.10\pm 0.06$&$20.0\pm 1.7$&$ 19.48\pm0.20$&$ 0.015\pm 0.009$&$ -6.14\pm0.35$ \cr
2016-12a & Silver & 7735.359&$ 17.20\pm0.08$&$ -7.36\pm0.13$&$41.4\pm 9.0$&$ 1.62\pm 0.09$&$34.1\pm 4.5$&$ 19.27\pm0.17$&$ 0.007\pm 0.017$&$ -6.64\pm0.19$ \cr
2016-12c & silver & 7745.533&$ 16.73\pm0.08$&$ -7.83\pm0.13$&$22.3\pm 6.0$&$ 1.35\pm 0.12$&$20.8\pm 3.6$&$ 18.74\pm0.24$&$ 0.020\pm 0.020$&$ -6.49\pm0.37$ \cr
2017-01d & Gold & 7775.199&$ 16.30\pm  0.10$&$ -8.26\pm  0.14$&$  19.1\pm   9.7$&$  1.28\pm  0.22$&$  10.3\pm   2.6$&$ 17.90\pm  0.29$&$ 0.045\pm 0.033$&$ -6.44\pm  0.36$ \cr
2017-11e & Silver & 8074.282&$ 16.40\pm0.10$&$ -8.16\pm0.14$&$ 4.7\pm 1.2$&$ 0.67\pm 0.11$&$ 7.4\pm 1.1$&$ 19.53\pm0.35$&$ 0.041\pm 0.042$&$ -4.72\pm0.49$ \cr
\enddata
\end{deluxetable*}

\begin{figure*}
\plotone{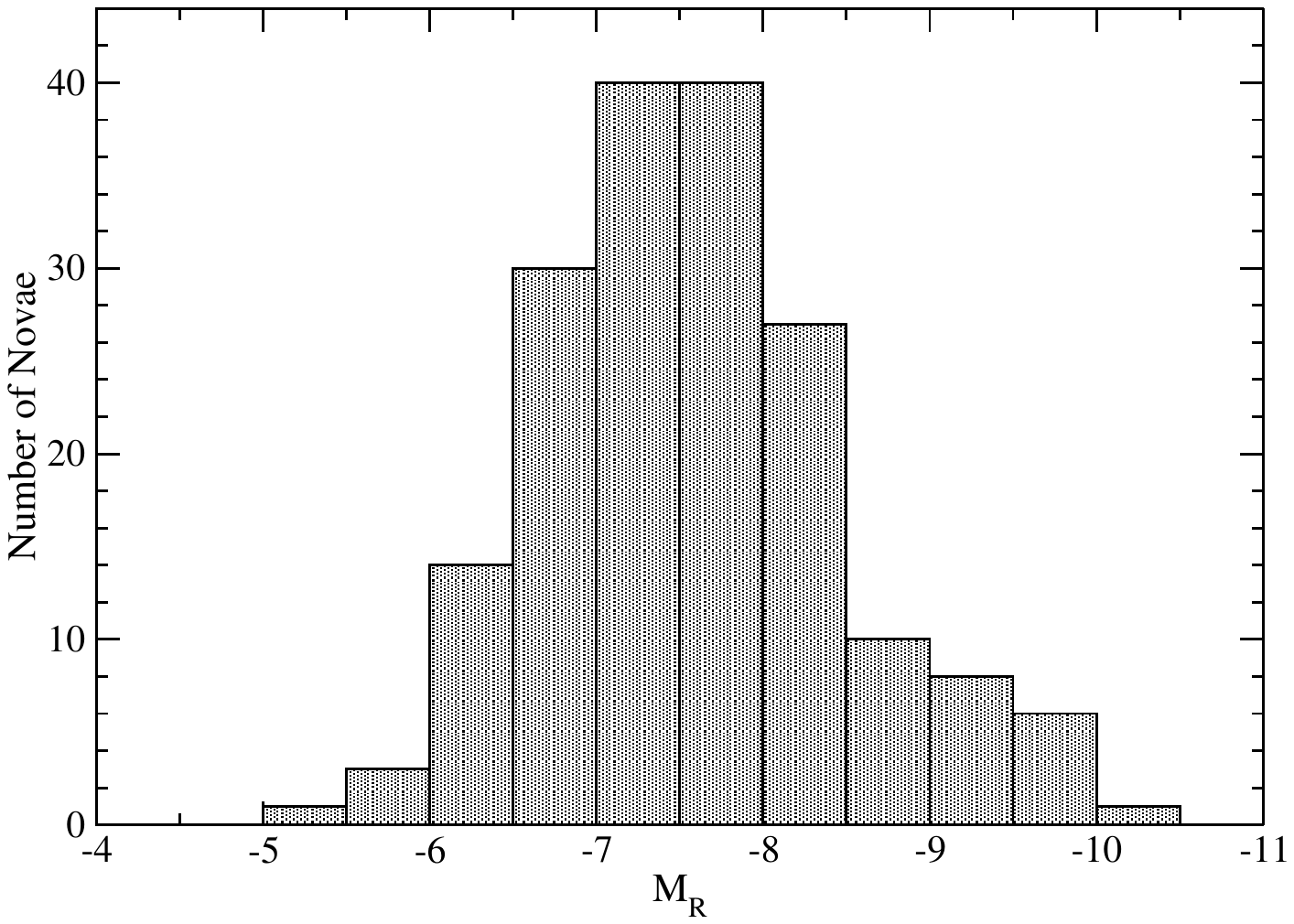}
\caption{The $R$-band $M_R$ distribution for our sample of M31 novae.
The dispersion of the distribution is characterized by a standard deviation $\sigma=0.89$~mag, with
an average given by
$\langle M_R \rangle=-7.57\pm0.07$.
The mean of the distribution is in excellent agreement with the recent mean for Galactic novae found by \citet{2022MNRAS.517.6150S}.
}
\label{fig:f4}
\end{figure*}

\begin{deluxetable*}{lrcccrcc}
\tablenum{3}
\tablecolumns{8}
\tabletypesize{\scriptsize}
\tablecaption{Jitter Nova Light Curve Parameters\label{tab3}}
\tablehead{\colhead{Nova} & & \colhead{JD (max)} &  &  & \colhead{$t_2$} & & \\ \colhead{(M31N)} & \colhead{Sample} & \colhead{(2,450,000+)} & \colhead{$m_R$ (max)} & \colhead{$M_R$ (max)} & \colhead{(d)} & \colhead{log $t_2$} & \colhead{$M_{15}$}}
\startdata
2006-10a& Bronze & 4043.331 & $  17.90\pm 0.10$&$  -6.66\pm 0.14$&$  67.6\pm  34.4$& $ 1.83\pm 0.22$&$-6.22\pm 0.27$ \cr
2008-07a& Bronze & 4620.542 & $  18.70\pm 0.25$&$  -5.86\pm 0.27$&$ 482.5\pm 103.5$& $ 2.68\pm 0.09$&$-5.80\pm 0.27$ \cr
2009-08a& Bronze & 5076.478 & $  17.80\pm 0.10$&$  -6.76\pm 0.14$&$  92.0\pm  36.8$& $ 1.96\pm 0.17$&$-6.43\pm 0.19$ \cr
2010-06b& Bronze & 5380.497 & $  17.65\pm0.10$&$  -6.91\pm0.14$&$  69.7\pm  62.6$& $1.84\pm0.39$&$  -6.48\pm0.41$ \cr
2010-06d& Bronze & 5382.456 & $  17.68\pm0.10$&$  -6.88\pm0.14$&$  95.0\pm  91.5$& $1.98\pm0.42$&$  -6.56\pm0.34$ \cr
2010-10b& Bronze & 5505.338 & $  17.60\pm0.10$&$  -6.96\pm0.14$&$ 211.4\pm 151.1$& $2.33\pm0.31$&$  -6.82\pm0.17$ \cr
2010-11a& Bronze & 5561.213 & $  18.10\pm0.10$&$  -6.46\pm0.14$&$  44.9\pm  25.6$& $1.65\pm0.25$&$  -5.79\pm0.41$ \cr
2011-05a& Bronze & 5718.549 & $  17.00\pm0.20$&$  -7.56\pm0.22$&$  74.3\pm  25.2$& $1.87\pm0.15$&$  -7.16\pm0.26$ \cr
2011-11a& Silver & 5885.214 & $  17.75\pm0.10$&$  -6.81\pm0.14$&$  33.2\pm   8.9$& $1.52\pm0.12$&$  -5.91\pm0.28$ \cr
2012-05b& Bronze & 6052.579 & $  18.20\pm0.15$&$  -6.36\pm0.18$&$  60.2\pm  28.0$& $1.78\pm0.20$&$  -5.86\pm0.29$ \cr
2013-10b& Bronze & 6574.234 & $  17.40\pm0.15$&$  -7.16\pm0.18$&$  27.1\pm   7.5$& $1.43\pm0.12$&$  -6.05\pm0.36$ \cr
2013-12a& Bronze & 6643.287 & $  18.10\pm0.10$&$  -6.46\pm0.14$&$ 166.8\pm  58.9$& $2.22\pm0.15$&$  -6.28\pm0.16$ \cr
2014-05b& Bronze & 6809.552 & $  17.50\pm0.15$&$  -7.06\pm0.18$&$  54.4\pm  23.1$& $1.74\pm0.18$&$  -6.51\pm0.30$ \cr
2014-10a& Bronze & 6944.663 & $  17.69\pm0.07$&$  -6.87\pm0.12$&$  91.1\pm  55.0$& $1.96\pm0.26$&$  -6.54\pm0.23$ \cr
2014-11a& Bronze & 6974.190 & $  17.70\pm0.15$&$  -6.86\pm0.18$&$ 110.2\pm  32.1$& $2.04\pm0.13$&$  -6.59\pm0.20$ \cr
2014-12b& Bronze & 7007.181 & $  17.50\pm0.10$&$  -7.06\pm0.14$&$  38.1\pm  32.1$& $1.58\pm0.37$&$  -6.27\pm0.68$ \cr
2015-03a& Bronze & 7208.546 & $  18.00\pm0.10$&$  -6.56\pm0.14$&$ 127.1\pm  72.0$& $2.10\pm0.25$&$  -6.32\pm0.19$ \cr
2015-04a& Bronze & 7207.564 & $  17.80\pm0.10$&$  -6.76\pm0.14$&$ 116.5\pm  49.0$& $2.07\pm0.18$&$  -6.50\pm0.18$ \cr
2015-05c& Bronze & 7179.524 & $  18.10\pm0.15$&$  -6.46\pm0.18$&$  82.2\pm  24.2$& $1.91\pm0.13$&$  -6.09\pm0.21$ \cr
2015-06b& Bronze & 7199.520 & $  17.40\pm0.10$&$  -7.16\pm0.14$&$  87.2\pm  72.6$& $1.94\pm0.36$&$  -6.82\pm0.32$ \cr
2015-10b& Bronze & 7360.304 & $  17.31\pm0.09$&$  -7.25\pm0.13$&$ 100.3\pm  75.8$& $2.00\pm0.33$&$  -6.95\pm0.26$ \cr
2015-11c& Bronze & 7350.546 & $  17.40\pm0.10$&$  -7.16\pm0.14$&$  51.8\pm  18.2$& $1.71\pm0.15$&$  -6.58\pm0.25$ \cr
2016-03d& Bronze & 7465.268 & $  17.60\pm0.15$&$  -6.96\pm0.18$&$  80.1\pm  24.5$& $1.90\pm0.13$&$  -6.59\pm0.21$ \cr
2016-08e& Silver & 7633.330 & $  17.10\pm0.10$&$  -7.46\pm0.14$&$  60.6\pm  16.7$& $1.78\pm0.12$&$  -6.97\pm0.20$ \cr
2016-09a& Bronze & 7658.256 & $  17.50\pm0.10$&$  -7.06\pm0.14$&$  57.0\pm  35.1$& $1.76\pm0.27$&$  -6.53\pm0.35$ \cr
2016-09b& Silver & 7660.325 & $  17.30\pm0.10$&$  -7.26\pm0.14$&$  43.9\pm  17.3$& $1.64\pm0.17$&$  -6.58\pm0.30$ \cr
2016-10a& Silver & 7694.346 & $  17.70\pm0.10$&$  -6.86\pm0.14$&$  35.5\pm  13.0$& $1.55\pm0.16$&$  -6.02\pm0.34$ \cr
\enddata
\end{deluxetable*}

\begin{deluxetable*}{lrcclrcc}
\tablenum{4}
\tablecolumns{8}
\tabletypesize{\scriptsize}
\tablecaption{Recurrent Nova Light Curve Parameters\label{tab4}}
\tablehead{\colhead{Nova} & & \colhead{JD (max)} &  &  & \colhead{$t_2$} & & \\ \colhead{(M31N)} & \colhead{Sample} & \colhead{(2,450,000+)} & \colhead{$m_R$ (max)} & \colhead{$M_R$ (max)} & \colhead{(d)} & \colhead{log $t_2$} & \colhead{$M_{15}$}}
\startdata
2007-08d & Silver & 4380.424 & $ 18.50\pm 0.15$&$ -6.06\pm 0.18$&$  13.0\pm 1.5$& $ 1.12\pm 0.05$&$ -3.76\pm 0.33$ \cr
2007-10b & Bronze & 4387.561 & $ 18.20\pm 0.10$&$ -6.36\pm 0.14$&$   2.8\pm 0.6$& $ 0.44\pm 0.10$&$\dots$ \cr
2008-12a\tablenotemark{a} & \dots & \dots & \dots & $-6.25\pm0.04$ & $2.2\pm0.1$  & $0.34\pm0.02$ & \dots \cr
2009-11b & Silver & 5135.632 & $ 18.60\pm 0.10$&$ -5.96\pm 0.14$&$ 110.4\pm29.2$& $ 2.04\pm 0.11$&$ -5.69\pm 0.16$ \cr
2010-10e & Bronze & 5501.427 & $  18.05\pm0.10$&$  -6.51\pm0.14$&$   4.3\pm0.5$& $0.63\pm0.05$&$   \dots$ \cr
2012-01b & Silver & 5950.233 & $  17.60\pm0.15$&$  -6.96\pm0.18$&$  12.6\pm0.9$& $1.10\pm0.03$&$  -4.58\pm0.25$ \cr
2012-09a & Bronze & 6179.270 & $  16.30\pm0.15$&$  -8.26\pm0.18$&$   9.7\pm1.4$& $0.99\pm0.06$&$  -5.17\pm0.47$ \cr
2013-05b & Bronze & 6441.533 & $  17.80\pm0.10$&$  -6.76\pm0.14$&$   4.7\pm0.2$& $0.67\pm0.02$&$   \dots $ \cr
2013-10c\tablenotemark{b} & Gold & 6579.218 & $  15.90\pm0.10$&$  -8.66\pm0.14$&$   5.5\pm1.7$& $0.74\pm0.13$&$  -4.66\pm  0.51$ \cr
2015-02b & Bronze & 7073.338 & $  16.54\pm0.05$&$  -8.02\pm0.11$&$   1.8\pm0.5$& $0.26\pm0.12$&$   \dots $ \cr
2015-10c & Bronze & 7327.241 & $  17.85\pm0.10$&$  -6.71\pm0.14$&$   3.6\pm0.5$& $0.55\pm0.07$&$   \dots $ \cr
2016-07e & Silver & 7598.355 & $  17.56\pm0.10$&$  -7.00\pm0.14$&$  11.1\pm1.7$& $1.04\pm0.07$&$  -4.30\pm0.45$ \cr
2016-12e & Bronze & 7751.236 & $  17.34\pm0.08$&$  -7.22\pm0.13$&$  10.2\pm2.6$& $1.01\pm0.11$&$  -4.29\pm0.77$ \cr
2017-12a & Bronze & 8113.176 & $  18.50\pm0.15$&$  -6.06\pm0.18$&$   2.7\pm0.1$& $0.43\pm0.02$&$   \dots $ \cr
2020-01b & Gold & 8878.234 & $  17.40\pm0.15$&$  -7.16\pm0.18$&$  11.4\pm2.5$& $1.06\pm0.09$&$  -4.53\pm0.60$ \cr
2022-09a & Gold & 9831.593 & $  17.70\pm0.10$&$  -6.86\pm0.14$&$  10.8\pm1.4$& $1.03\pm0.06$&$  -4.08\pm0.40$ \cr
2022-11b & Silver & 9896.208 & $  16.13\pm0.07$&$  -8.43\pm0.12$&$   9.8\pm1.6$& $0.99\pm0.07$&$  -5.37\pm0.52$ \cr
\enddata
\tablenotetext{a}{Data from \citet{2023RNAAS...7..191S}.}
\tablenotetext{b}{Newly-identified RN \citep[see][for determination of parameter values]{2024RNAAS...8....5S}.}
\end{deluxetable*}

\begin{deluxetable*}{lrcrc}
\tablenum{5}
\tablecolumns{5}
\tabletypesize{\scriptsize}
\tablecaption{Predicted Recurrent Novae\label{tab5}}
\tablehead{\colhead{Nova} & & & \colhead{$t_2$} & \\ \colhead{(M31N)} & \colhead{Sample} & \colhead{$M_R$ (max)} & \colhead{(d)} & \colhead{$M_{15}$}}
\startdata
2005-07a & Bronze & $-7.16\pm0.18$&$   9.4\pm6.1$ &$ -3.95\pm2.11$ \cr
2007-12b & Bronze & $-7.56\pm0.14$&$   6.5\pm0.5$ &$ -2.93\pm0.37$ \cr
2008-07b & Silver & $ -6.16\pm0.14$&$  23.5\pm7.7$ &$ -4.89\pm0.44$ \cr
2008-11a & Bronze & $ -8.06\pm0.14$&$   4.8\pm1.5$ &$ -1.81\pm1.93$ \cr
2010-09b & Gold   & $ -6.36\pm0.18$&$  18.0\pm2.2$ &$ -4.69\pm0.27$ \cr
2011-01b & Gold   & $ -8.20\pm0.11$&$   3.4\pm0.9$ &$ -2.75\pm1.43$ \cr
2011-02b & Bronze & $ -6.86\pm0.22$&$  14.0\pm2.6$ &$ -4.72\pm0.45$ \cr
2012-06e & Gold   & $ -6.56\pm0.18$&$  16.1\pm12.1$&$ -4.62\pm0.64$ \cr
2013-02b & Bronze & $ -6.76\pm0.14$&$  14.9\pm1.1$ &$ -4.75\pm0.21$ \cr
2013-08a & Gold   & $ -6.96\pm0.18$&$   8.7\pm1.4$ &$ -3.49\pm0.60$ \cr
2015-11b & Bronze & $ -5.36\pm0.18$&$  32.2\pm5.5$ &$ -4.43\pm0.24$ \cr
2017-02a & Bronze & $ -7.26\pm0.14$&$   6.7\pm0.7$ &$ -2.80\pm0.51$ \cr
\enddata
\end{deluxetable*}

\begin{figure*}
\plotone{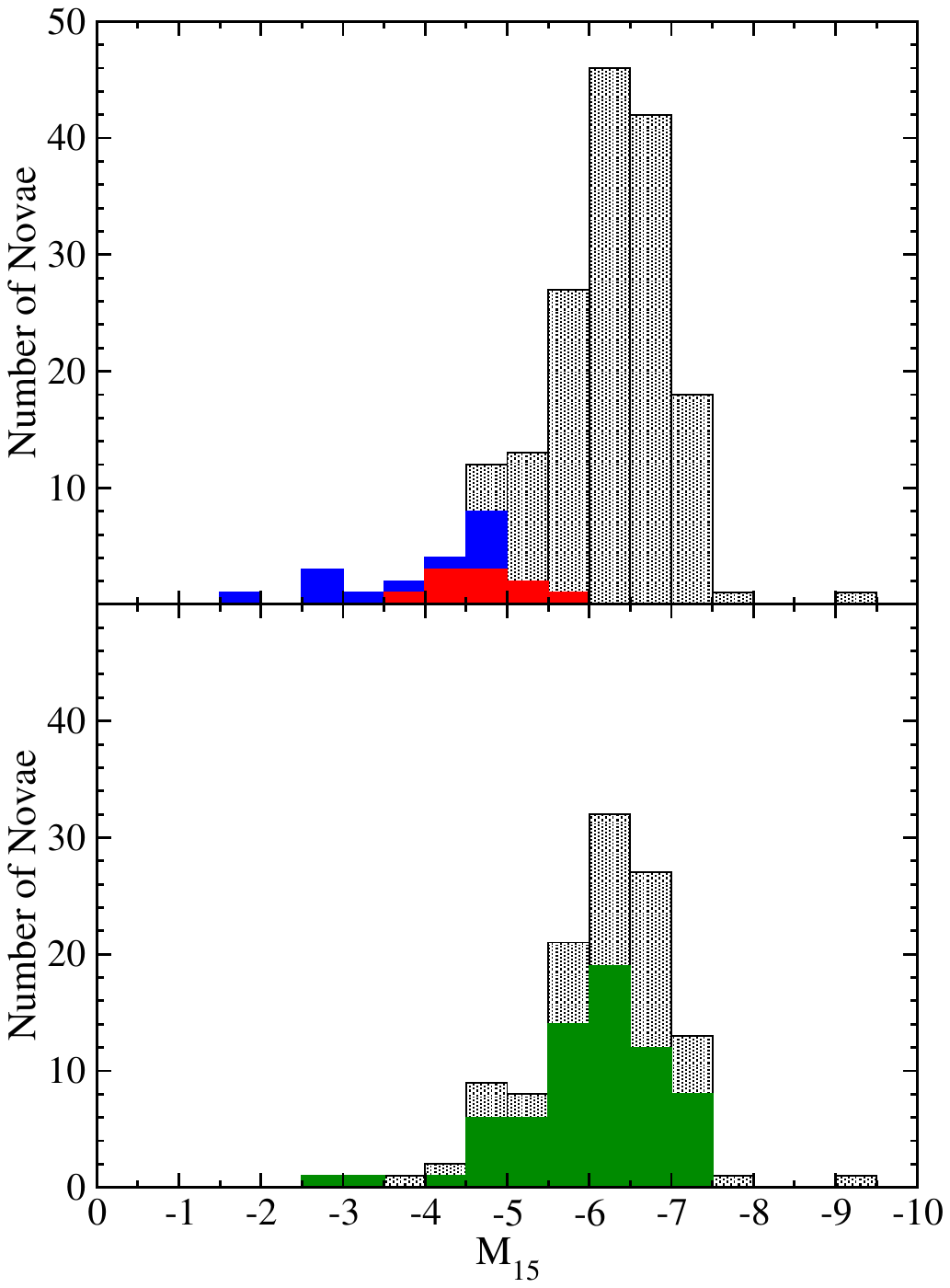}
\caption{{\it Top panel:} The $R$-band $M_{15}$ distribution for our complete sample of M31 novae. The red region shows the $M_{15}$ values
for the known recurrent novae, while the blue region shows the suspected recurrent nova systems.
The long tail to fainter magnitudes
results primarily from the extrapolation of the rate of decline of novae in the second quadrant of the MMRD relation.
{\it Bottom panel:} The $R$-band $M_{15}$ distribution for our high quality samples of M31 novae. The lightly
shaded region shows the novae in our silver sample, while the dark green region
shows novae with the best light curves in our gold sample.
}
\label{fig:f5}
\end{figure*}

\subsection{Comparison with the Galactic MMRD Relation}

The MMRD relation for Galactic novae has been studied by several authors over the past 50 years \citep[e.g.,][]{1985ApJ...292...90C,2000AJ....120.2007D,2018MNRAS.476.4162O,2022MNRAS.517.6150S}. The earlier studies of \citet{1985ApJ...292...90C} and \citet{2000AJ....120.2007D} relied on relatively small
samples of novae with distances based primarily on expansion parallaxes. Both of these studies found significant correlations
of absolute magnitude with rate of decline with MMRD relations of
$M_V = -10.7~(\pm0.3) + 2.41~(\pm0.23)~\log t_2$ and $M_V = -11.32~(\pm0.44) + 2.55~(\pm0.32)~\log t_2$
for the \cite{1985ApJ...292...90C} and \citet{2000AJ....120.2007D} studies, respectively. More recently, \citet{2016MNRAS.461.1177O} determined reddening-distance relations toward a large sample of Galactic novae ($N=119$), and then used these relations to estimate distances to 73 Galactic novae. Subsequently, \citet{2018MNRAS.476.4162O} supplemented the most reliable of these distances with expansion parallax determinations to explore a variety of Galactic MMRD relations based on different sub-samples of novae. Their overall relation, based on a sample of 50 novae
with the most reliable distances produced an MMRD relation given by $M_V = -10.54~(\pm0.30) + 2.04~(\pm0.23)~\log t_2$.
The scatter in their final relation, $\sigma=0.9$~mag, is comparable to what we find for our M31 MMRD relation. 
The most recent exploration of the MMRD relation for Galactic novae has been undertaken by \citet{2022MNRAS.517.6150S}. As we have seen for M31, Schaefer found a weak correlation ($M_V\simeq-9.1+1.5~\log t_3$) between the peak luminosity of a sample of 192 Galactic novae and their rates of decline. However, given the large scatter about this relation ($\sigma\simeq1$~mag) he concluded that it was of no value for distance determinations.

In all cases we have considered the slope of the Galactic MMRD relation is somewhat steeper than that found for our M31 nova sample, although given the uncertainties it is unclear whether this discrepancy reflects any real difference between the Galactic and M31 nova populations. The principal take away from these studies is that, for both M31 and the Galaxy,
the significant scatter renders the MMRD relation practically useless for determining reliable extragalactic distances.
However, as we will explore in section~\ref{sec:M15} below, making use of the nova luminosity function 15 days after maximum light may offer a more promising avenue for estimating extragalactic distances.

\section{The Nova Luminosity Distribution at Maximum Light} \label{sec:LDmax}

Figure~\ref{fig:f4} shows the $R$-band luminosity function for M31 novae at maximum light. The distribution of peak absolute magnitudes is approximately Gaussian, and is characterized by a standard deviation $\sigma=0.89$~mag and a mean of $\langle M_{R} \rangle = -7.57\pm0.07$.
For comparison, \citet{2009ApJ...690.1148S} estimated a $V$-band luminosity distribution for the 732 novae with photometric measurements recorded in M31 up through 2007 finding an average absolute magnitude $\langle M_V \rangle = -7.20\pm0.04$.
A part of the difference between this value and the mean absolute magnitude
of our $R$-band sample can be attributed to the difference in bandpass ($V$ vs. $R$). Adopting $(V-R)=0.16$ as a reasonable estimate of the color of a nova near maximum light \citep[][]{2009ApJ...690.1148S} would suggest $\langle M_R \rangle \simeq -7.36$ for the earlier sample. The rest of the discrepancy can likely be attributed to the
fact that complete light curves were unavailable for most of the novae used in the \citet{2009ApJ...690.1148S} analysis, and thus maximum light was more likely to be missed compared with the present study.

In the study of Galactic novae by \citet{2022MNRAS.517.6150S} mentioned earlier,
a mean absolute magnitude, $\langle M_V \rangle = -7.45\pm0.09$\footnote{We have converted the standard deviation for the full distribution ($\sigma=1.33$) to a standard deviation of the mean ($\sigma_m=1.33/\sqrt{213}=0.09$).} was found based
on his complete sample of 213 novae with available distances and peak magnitudes.
Correcting for the approximate $V-R$ color at maximum light suggests a mean $R$-band absolute magnitude distribution
for Galactic novae of $\langle M_R \rangle = -7.61\pm0.09$, which agrees extremely well with the mean of our M31 $R$-band sample. If a suitably large sample of novae were available in a given galaxy, a comparison of the absolute magnitude
distribution at maximum light with that found in M31 or the Galaxy could in principle be used to estimate distances to an accuracy of perhaps $\sim15$\%.

\section{The $M_{15}$ Relation} \label{sec:M15}

A consequence of an MMRD relation where bright novae generally fade more quickly than their less luminous counterparts is that the light curves of the novae must intersect -- that is they must reach the same luminosity -- at some point in their post maximum evolution. Thus, the determination of a nova luminosity distribution 15 days after maximum light might be expected to
provide a better determination of extragalactic distances than could be achieved by using the luminosity distribution
at maximum light.
In a study of novae in the Magellanic Clouds, \citet{1955Obs....75..170B} showed that novae tended to reach a common luminosity at $\sim$15~days post
maximum light, finding that $M_{15,pg} = -5.2\pm0.1$ (p.e.). In recent years
recalibrations of the $M_{15}$ relation for Galactic novae given by \citet{1985ApJ...292...90C} and \citet{2000AJ....120.2007D} who found $M_{15,V} = -5.60\pm0.45$ and $M_{15,V} = -6.05\pm0.44$, respectively.

We have measured $M_{15,R}$ values for the large sample of novae in our M31 data\footnote{In a few cases among the recurrent novae, the light curves did not extend
sufficiently far after maximum light for a meaningful $M_{15}$ value to be estimated.}, and plotted the distribution in the upper panel of Figure~\ref{fig:f5}.
The overall distribution is peaked near $M_{15,R} \sim -6$, albeit with an extended tail towards fainter luminosity.
Assuming the distribution to be Gaussian, we find it to be characterized by a standard deviation $\sigma=1.0$~mag with a mean given by $\langle M_{15,R} \rangle = -6.06\pm0.08$.
When known recurrent novae are excluded, the distribution shifts slightly to higher luminosity and tightens a bit as expected giving
$\langle M_{15,R} \rangle = -6.15\pm0.07$ distributed with a standard deviation, $\sigma=0.95$~mag. Overall the mean of our $M_{15,R}$ distribution is in good
agreement with the mean $M_{15,V}$ value found by Downes \& Duerbeck for their sample of 28 Galactic novae, however our distribution is significantly broader.
Indeed, contrary to our expectations, the dispersion in our $M_{15,R}$ distribution for M31 novae is comparable to that seen
in the $M_R$ distribution at maximum light suggesting that the characteristic luminosity $\sim$15 days after maximum light may not converge
to a common value for all novae.

In an attempt to reduce the dispersion in the $M_{15}$ distribution, we have explored the effect of restricting our $M_{15,R}$ sample to novae with the highest quality light curves. The bottom panel of Figure~\ref{fig:f5} shows the $M_{15,R}$ distributions for the 67 novae in our gold sample and the 118 novae in the silver sample.
The distributions are characterized by means
of $\langle M_{15,R} \rangle = -5.99\pm0.11$ and $\langle M_{15,R} \rangle = -6.13\pm0.08$, and standard
deviations of $\sigma=0.90$ and $\sigma=0.91$~mag for the gold and silver samples, respectively.
As we discovered with the MMRD relation, the dispersion of the $M_{15,R}$ distribution for our high-quality sample of M31 novae does not appear to provide a significant improvement over that for the full nova sample.

For both the full and the restricted samples much of the dispersion in the $M_{15}$
distribution comes from its extended left tail. If we restrict our analysis to novae brighter than
$M_R=-5$ (i.e., we exclude all known and suspected recurrent novae), the resulting symmetrical peak region
of the full distribution yields $\langle M_{15,R} \rangle = -6.36\pm0.05$ with a reduced standard deviation
of just $\sigma=0.62$~mag.
Thus, by restricting the analysis to the symmetrical portion of the distribution, it appears that use of the $M_{15}$ distribution may offer a modest advantage over the $M_R$ distribution at maximum light for extragalactic
distance estimates. That said, given that determining the absolute magnitude distribution at maximum light requires significantly less observational effort, use of the $M_{15}$ distribution for extragalactic distance determinations may not be warranted in all situations.

In addition to perhaps offering a modest advantage over the $M_\mathrm{max}$ distribution for distance determinations, the asymmetric nature of the $M_{15}$ distribution
may suggest an effective way to identify novae that are potentially recurrent. Known recurrent novae for which we were able to
determine $M_{15,R}$ values are indicated by the red histogram in the top panel of Figure~\ref{fig:f5}.
All such systems lie in the extended low luminosity tail of the overall distribution
as expected for a population of faint and fast novae falling in the lower left quadrant (Q2) of the MMRD relation.
We also have highlighted in the blue histogram other novae from Q2 of the MMRD plot that have not (yet) been observed to be recurrent. These systems, which we
have compiled in Table~\ref{tab5}, represent ``faint and fast" novae that mostly lie in the lower left quadrant of the MMRD plot below the lower dotted line in Figure~\ref{fig:f2}. We predict that these systems, which
should have relatively short recurrence times,
are likely to be recognized as recurrent novae in the relatively near future ($\lessim 100$~yrs).

\section{Conclusions} \label{sec:Conclusions}

We have presented the analysis of a large and homogeneous sample of $R$-band light curves of novae in M31. The principal conclusions of our study
can be summarized as follows:

(1) We find an approximate correlation between the peak luminosity of a nova and its rate of decline from maximum light consistent with
a canonical MMRD relation. The scatter in the relationship is large, however, with known recurrent novae departing significantly from
the best-fit MMRD relation. The weakness of the correlation of $M_R$ with $\log t_2$ renders the MMRD essentially useless for deriving the
distances to individual novae with measured light curves.

(2) The slowest declining novae typically have erratic declines from peak luminosity and comprise virtually
all off the Jitter class novae.

(3) The overall spread of nova luminosities span over a factor of $\sim$100, and are characterized by absolute $R$-band magnitudes ranging from
$M_R\simeq-5.5$ to $M_R\simeq-10.5$. The novae are distributed with standard deviation $\sigma=0.89$~mag about a mean $R$-band absolute magnitude
given by $\langle M_R \rangle=-7.57\pm0.07$.
When known recurrent novae are excluded,
the mean absolute magnitude is slightly brighter, and is given by $\langle M_R \rangle=-7.61\pm0.07$. The overall M31 luminosity distribution
is in excellent agreement with that found by \citet{2022MNRAS.517.6150S} for Galactic novae suggesting that the nova populations in M31
and the Galaxy are quite similar.

(4) The distribution of nova $R$-band absolute magnitudes 15 days after peak luminosity, $M_{15}(R)$, is characterized by 
a standard deviation $\sigma\simeq1$~mag centered on a mean given by $\langle M_{15}\rangle=-6.06\pm0.08$.
Surprisingly, the dispersion in the
$M_{15}$ relation appears to be comparable to
that for the luminosity distribution at maximum light.
However, if we restrict the analysis to the symmetrical portion of the $M_{15}$ distribution, neglecting novae
on the faint tail with $M_R\leq-5$, we find a distribution mean of $\langle M_{15}\rangle=-6.36\pm0.05$ with
a reduced standard deviation of just $\sigma=0.62$~mag.
Thus, the truncated $M_{15}$ relation may offer a modest advantage to the use of the
luminosity distribution at maximum light for measuring extragalactic distances.
The determination of a useful distance modulus for an external galaxy
relative to that of M31 good to $\sim0.1$~mag would require a (perhaps impractically)
large ($\grtsim50$) sample of novae light curves.

(5) By studying the MMRD relation and the $M_{15}$ distribution, we have identified a dozen faint and fast
M31 novae that likely have recurrence times $\lessim100$~years. Future
observations should reveal many of these systems to be members of the recurrent novae population of M31.

Our overarching conclusion based on our extensive sample of $R$-band light curves is that the peak luminosities of M31 novae are only weakly correlated with their rates of decline. As has been found in recent studies of Galactic novae, the resulting MMRD relation is found to be poorly suited for use in determining extragalactic distances.
Furthermore, despite earlier results suggesting that the luminosity of novae 15~days after maximum light, $M_{15}$, might approach a characteristic value useful for distance
determinations, we have found no evidence that the dispersion in $M_{15}$ for our $R$-band sample of M31 novae
is significantly smaller that their dispersion at maximum light.

\begin{acknowledgments}
We would like to thank the following observers who have taken supplementary M31 images used for
the photometry: P. Hornochov\'a, P. Zasche, P. Kub\'anek, J. Gorosabel, K. Ho\v{n}kov\'a,
O. Vaduvescu, M. Jel\'inek, B. Mikuleck\'a, A. Manilla-Robles, A. Valeev, S. Mottola,
G. Ramsay, S. Hellmich, V. Tudor, Z. Jan\'ak, P. Garnavich, J. Casares, J. Corral-Santana,
A. Gal\'ad, R. Khan, C. Zurita, P. Scheirich, O. Sholukhova, S. Vinokurov, J. Pagnini,
A. Kaur, P. Fatka, M. Orio, J. Vilagi, M. Lehk\'y, O. Lara Gil, D. Pinfield, M. Velen,
M. Zejda, N. Paul, B. Sipocz, G. Kovacs, C. Littlefield, A. Gonzalez, S. Bouzid, M. Skarka,
N. Morales, L. \v{R}ezba, J. Bird, R. Hueso, S. Perez-Hoyos, J. Prieto, A. Sanchez-Lavega,
O. Pejcha, A. Applegate, S. Kaisin, K. Magno, E. Barsukova, J. Gallagher, T. Fatkhullin,
N. Sarkisyan, P. Caga\v{s}, C. Kennedy, and F. Valeev.
Work of K.H. and P.K. was supported by the project RVO:67985815. This work is based (in part)
on data collected with the Danish 1.54-m telescope at the ESO La Silla Observatory.
\end{acknowledgments}

\vspace{5mm}
\facilities{0.65-m, Ond\v{r}ejov Observatory; Danish 1.54-m, La Silla Observatory; INT 2.54-m, Roque de los
Muchachos Observatory; CAHA 1.23-m, Calar Alto Astronomical Observatory; VATT 1.83-m, Mount
Graham International Observatory; SAO 6-m, Special Astrophysical Observatory RAS; Hiltner 2.4-m,
MDM Observatory; BOOTES-2 0.60-m, EELM-CSIC Boyden Observatory; 0.91-m, Kitt Peak National
Observatory; WIYN 3.5-m, Kitt Peak National Observatory; Blue Eye 600 0.60-m, Charles University;
0.30-m, Zl\'in Observatory.}

\newpage

\bibliography{novarefs}{}

\begin{thebibliography}{}
\expandafter\ifx\csname natexlab\endcsname\relax\def\natexlab#1{#1}\fi
\providecommand{\url}[1]{\href{#1}{#1}}
\providecommand{\dodoi}[1]{doi:~\href{http://doi.org/#1}{\nolinkurl{#1}}}
\providecommand{\doeprint}[1]{\href{http://ascl.net/#1}{\nolinkurl{http://ascl.net/#1}}}
\providecommand{\doarXiv}[1]{\href{https://arxiv.org/abs/#1}{\nolinkurl{https://arxiv.org/abs/#1}}}

\bibitem[{{Arp}(1956)}]{1956AJ.....61...15A}
{Arp}, H.~C. 1956, \aj, 61, 15, \dodoi{10.1086/107284}

\bibitem[{{Bode} \& {Evans}(2008)}]{2008clno.book.....B}
{Bode}, M.~F., \& {Evans}, A. 2008, {Classical Novae}, Vol.~43

\bibitem[{{Buscombe} \& {de Vaucouleurs}(1955)}]{1955Obs....75..170B}
{Buscombe}, W., \& {de Vaucouleurs}, G. 1955, The Observatory, 75, 170

\bibitem[{{Capaccioli} {et~al.}(1989){Capaccioli}, {Della Valle}, {D'Onofrio},
  \& {Rosino}}]{1989AJ.....97.1622C}
{Capaccioli}, M., {Della Valle}, M., {D'Onofrio}, M., \& {Rosino}, L. 1989,
  \aj, 97, 1622, \dodoi{10.1086/115104}

\bibitem[{{Ciardullo} {et~al.}(1987){Ciardullo}, {Ford}, {Neill}, {Jacoby}, \&
  {Shafter}}]{1987ApJ...318..520C}
{Ciardullo}, R., {Ford}, H.~C., {Neill}, J.~D., {Jacoby}, G.~H., \& {Shafter},
  A.~W. 1987, \apj, 318, 520, \dodoi{10.1086/165388}

\bibitem[{{Cohen}(1985)}]{1985ApJ...292...90C}
{Cohen}, J.~G. 1985, \apj, 292, 90, \dodoi{10.1086/163135}

\bibitem[{{de Vaucouleurs}(1978)}]{1978ApJ...223..351D}
{de Vaucouleurs}, G. 1978, \apj, 223, 351, \dodoi{10.1086/156269}

\bibitem[{{Della Valle} \& {Izzo}(2020)}]{2020A&ARv..28....3D}
{Della Valle}, M., \& {Izzo}, L. 2020, \aapr, 28, 3,
  \dodoi{10.1007/s00159-020-0124-6}

\bibitem[{{della Valle} {et~al.}(1994){della Valle}, {Rosino}, {Bianchini}, \&
  {Livio}}]{1994A&A...287..403D}
{della Valle}, M., {Rosino}, L., {Bianchini}, A., \& {Livio}, M. 1994, \aap,
  287, 403

\bibitem[{{Downes} \& {Duerbeck}(2000)}]{2000AJ....120.2007D}
{Downes}, R.~A., \& {Duerbeck}, H.~W. 2000, \aj, 120, 2007,
  \dodoi{10.1086/301551}

\bibitem[{{Gerasimovic}(1936)}]{1936PA.....44...78G}
{Gerasimovic}, B.~P. 1936, Popular Astronomy, 44, 78

\bibitem[{{Helton} {et~al.}(2014){Helton}, {Evans}, {Woodward}, {Gehrz}, \&
  {Vacca}}]{2014ASPC..490..261H}
{Helton}, L.~A., {Evans}, A., {Woodward}, C.~E., {Gehrz}, R.~D., \& {Vacca}, W.
  2014, in Astronomical Society of the Pacific Conference Series, Vol. 490,
  Stellar Novae: Past and Future Decades, ed. P.~A. {Woudt} \& V.~A.~R.~M.
  {Ribeiro}, 261

\bibitem[{{Hubble}(1929)}]{1929ApJ....69..103H}
{Hubble}, E.~P. 1929, \apj, 69, 103, \dodoi{10.1086/143167}

\bibitem[{{Kato} {et~al.}(2014){Kato}, {Saio}, {Hachisu}, \&
  {Nomoto}}]{2014ApJ...793..136K}
{Kato}, M., {Saio}, H., {Hachisu}, I., \& {Nomoto}, K. 2014, \apj, 793, 136,
  \dodoi{10.1088/0004-637X/793/2/136}

\bibitem[{{Massey} {et~al.}(2006){Massey}, {Olsen}, {Hodge}, {Strong},
  {Jacoby}, {Schlingman}, \& {Smith}}]{2006AJ....131.2478M}
{Massey}, P., {Olsen}, K.~A.~G., {Hodge}, P.~W., {et~al.} 2006, \aj, 131, 2478,
  \dodoi{10.1086/503256}

\bibitem[{{McLaughlin}(1939)}]{1939PA.....47..410M}
{McLaughlin}, D.~B. 1939, Popular Astronomy, 47, 410

\bibitem[{{McLaughlin}(1942)}]{1942PA.....50..233M}
---. 1942, Popular Astronomy, 50, 233

\bibitem[{{Mclaughlin}(1945)}]{1945PASP...57...69M}
{Mclaughlin}, D.~B. 1945, \pasp, 57, 69, \dodoi{10.1086/125689}

\bibitem[{{McLaughlin}(1960)}]{1960stat.book..585M}
{McLaughlin}, D.~B. 1960, in Stellar atmospheres. Edited by Jesse Leonard
  Greenstein. Supported in part by the National Science Foundation. Published
  by the University of Chicago Press, 585

\bibitem[{{Nomoto}(1982)}]{1982ApJ...253..798N}
{Nomoto}, K. 1982, \apj, 253, 798, \dodoi{10.1086/159682}

\bibitem[{{{\"O}zd{\"o}nmez} {et~al.}(2018){{\"O}zd{\"o}nmez}, {Ege},
  {G{\"u}ver}, \& {Ak}}]{2018MNRAS.476.4162O}
{{\"O}zd{\"o}nmez}, A., {Ege}, E., {G{\"u}ver}, T., \& {Ak}, T. 2018, \mnras,
  476, 4162, \dodoi{10.1093/mnras/sty432}

\bibitem[{{{\"O}zd{\"o}nmez} {et~al.}(2016){{\"O}zd{\"o}nmez}, {G{\"u}ver},
  {Cabrera-Lavers}, \& {Ak}}]{2016MNRAS.461.1177O}
{{\"O}zd{\"o}nmez}, A., {G{\"u}ver}, T., {Cabrera-Lavers}, A., \& {Ak}, T.
  2016, \mnras, 461, 1177, \dodoi{10.1093/mnras/stw1362}

\bibitem[{{Pfau}(1976)}]{1976A&A....50..113P}
{Pfau}, W. 1976, \aap, 50, 113

\bibitem[{{Pravec} {et~al.}(1994){Pravec}, {Hudec}, {Sold{\'a}n}, {Sommer}, \&
  {Schenkl}}]{1994ExA.....5..375P}
{Pravec}, P., {Hudec}, R., {Sold{\'a}n}, J., {Sommer}, M., \& {Schenkl}, K.~H.
  1994, Experimental Astronomy, 5, 375, \dodoi{10.1007/BF01583708}

\bibitem[{{Pritchet} \& {van den Bergh}(1987)}]{1987ApJ...318..507P}
{Pritchet}, C.~J., \& {van den Bergh}, S. 1987, \apj, 318, 507,
  \dodoi{10.1086/165387}

\bibitem[{{Quimby} {et~al.}(2021){Quimby}, {Shafter}, \&
  {Corbett}}]{2021RNAAS...5..160Q}
{Quimby}, R.~M., {Shafter}, A.~W., \& {Corbett}, H. 2021, Research Notes of the
  American Astronomical Society, 5, 160, \dodoi{10.3847/2515-5172/ac14c0}

\bibitem[{{Rector} {et~al.}(2022){Rector}, {Shafter}, {Burris}, {Walentosky},
  {Viafore}, {Strom}, {Cool}, {Sola}, {Crayton}, {Pilachowski}, {Jacoby},
  {Corbett}, {Rene}, \& {Hernandez}}]{2022ApJ...936..117R}
{Rector}, T.~A., {Shafter}, A.~W., {Burris}, W.~A., {et~al.} 2022, \apj, 936,
  117, \dodoi{10.3847/1538-4357/ac87ad}

\bibitem[{{Rosino}(1964)}]{1964AnAp...27..498R}
{Rosino}, L. 1964, Annales d'Astrophysique, 27, 498

\bibitem[{{Rosino}(1973)}]{1973Ros}
---. 1973, \aaps, 9, 347

\bibitem[{{Schaefer}(2018)}]{2018MNRAS.481.3033S}
{Schaefer}, B.~E. 2018, \mnras, 481, 3033, \dodoi{10.1093/mnras/sty2388}

\bibitem[{{Schaefer}(2022)}]{2022MNRAS.517.6150S}
---. 2022, \mnras, 517, 6150, \dodoi{10.1093/mnras/stac2900}

\bibitem[{{Schlafly} \& {Finkbeiner}(2011)}]{2011ApJ...737..103S}
{Schlafly}, E.~F., \& {Finkbeiner}, D.~P. 2011, \apj, 737, 103,
  \dodoi{10.1088/0004-637X/737/2/103}

\bibitem[{{Shafter}(2019)}]{2019enhp.book.....S}
{Shafter}, A.~W. 2019, {Extragalactic Novae; A historical perspective},
  \dodoi{10.1088/2514-3433/ab2c63}

\bibitem[{{Shafter} {et~al.}(2023){Shafter}, {Clark}, \&
  {Hornoch}}]{2023RNAAS...7..191S}
{Shafter}, A.~W., {Clark}, J.~G., \& {Hornoch}, K. 2023, Research Notes of the
  American Astronomical Society, 7, 191, \dodoi{10.3847/2515-5172/acf5e8}

\bibitem[{{Shafter} \& {Irby}(2001)}]{2001ApJ...563..749S}
{Shafter}, A.~W., \& {Irby}, B.~K. 2001, \apj, 563, 749, \dodoi{10.1086/324044}

\bibitem[{{Shafter} {et~al.}(2009){Shafter}, {Rau}, {Quimby}, {Kasliwal},
  {Bode}, {Darnley}, \& {Misselt}}]{2009ApJ...690.1148S}
{Shafter}, A.~W., {Rau}, A., {Quimby}, R.~M., {et~al.} 2009, \apj, 690, 1148,
  \dodoi{10.1088/0004-637X/690/2/1148}

\bibitem[{{Shafter} {et~al.}(2011){Shafter}, {Darnley}, {Hornoch},
  {Filippenko}, {Bode}, {Ciardullo}, {Misselt}, {Hounsell}, {Chornock}, \&
  {Matheson}}]{2011ApJ...734...12S}
{Shafter}, A.~W., {Darnley}, M.~J., {Hornoch}, K., {et~al.} 2011, \apj, 734,
  12, \dodoi{10.1088/0004-637X/734/1/12}

\bibitem[{{Shafter} {et~al.}(2024){Shafter}, {Hornoch},
  {Ku{\v{c}}{\'a}kov{\'a}}, {Fatka}, {Zhao}, {Gao}, {Yaqup}, {Zhong},
  {Esamdin}, {Bai}, {Wang}, {Benni}, {Luo}, \& {Yousuf}}]{2024RNAAS...8....5S}
{Shafter}, A.~W., {Hornoch}, K., {Ku{\v{c}}{\'a}kov{\'a}}, H., {et~al.} 2024,
  Research Notes of the American Astronomical Society, 8, 5,
  \dodoi{10.3847/2515-5172/ad19de}

\bibitem[{{Simien} {et~al.}(1978){Simien}, {Athanassoula}, {Pellet}, {Monnet},
  {Maucherat}, \& {Court{\`e}s}}]{1978A&A....67...73S}
{Simien}, F., {Athanassoula}, E., {Pellet}, A., {et~al.} 1978, \aap, 67, 73

\bibitem[{{Skopal}(2008)}]{2008JAVSO..36....9S}
{Skopal}, A. 2008, \jaavso, 36, 9, \dodoi{10.48550/arXiv.0805.1222}

\bibitem[{{Starrfield} {et~al.}(2016){Starrfield}, {Iliadis}, \&
  {Hix}}]{2016PASP..128e1001S}
{Starrfield}, S., {Iliadis}, C., \& {Hix}, W.~R. 2016, \pasp, 128, 051001,
  \dodoi{10.1088/1538-3873/128/963/051001}

\bibitem[{{Strope} {et~al.}(2010){Strope}, {Schaefer}, \&
  {Henden}}]{2010AJ....140...34S}
{Strope}, R.~J., {Schaefer}, B.~E., \& {Henden}, A.~A. 2010, \aj, 140, 34,
  \dodoi{10.1088/0004-6256/140/1/34}

\bibitem[{{Townsley} \& {Bildsten}(2005)}]{2005ApJ...628..395T}
{Townsley}, D.~M., \& {Bildsten}, L. 2005, \apj, 628, 395,
  \dodoi{10.1086/430594}

\bibitem[{{Tully} {et~al.}(2013){Tully}, {Courtois}, {Dolphin}, {Fisher},
  {H{\'e}raudeau}, {Jacobs}, {Karachentsev}, {Makarov}, {Makarova},
  {Mitronova}, {Rizzi}, {Shaya}, {Sorce}, \& {Wu}}]{2013AJ....146...86T}
{Tully}, R.~B., {Courtois}, H.~M., {Dolphin}, A.~E., {et~al.} 2013, \aj, 146,
  86, \dodoi{10.1088/0004-6256/146/4/86}

\bibitem[{{van den Bergh}(1988)}]{1988ASPC....4..221V}
{van den Bergh}, S. 1988, in Astronomical Society of the Pacific Conference
  Series, Vol.~4, The Extragalactic Distance Scale, ed. S.~{van den Bergh} \&
  C.~J. {Pritchet}, 221--229

\bibitem[{{van den Bergh} \& {Pritchet}(1986)}]{1986PASP...98..110V}
{van den Bergh}, S., \& {Pritchet}, C.~J. 1986, \pasp, 98, 110,
  \dodoi{10.1086/131728}

\bibitem[{{Warner}(1995)}]{1995cvs..book.....W}
{Warner}, B. 1995, {Cataclysmic variable stars}, Vol.~28

\bibitem[{{Zwicky}(1936)}]{1936PASP...48..191Z}
{Zwicky}, F. 1936, \pasp, 48, 191, \dodoi{10.1086/124698}

\end{thebibliography}
\bibliographystyle{aasjournal}

\appendix
\startlongtable
\begin{deluxetable}{cccc}
\tabletypesize{\scriptsize}
\tablenum{A1}
\tablecolumns{4}
\tablecaption{Nova Photometry}
\tablehead{\colhead{JD$-2,450,000$} & \colhead{Filter} & \colhead{Magnitude} & \colhead{Error}
}
\startdata
\cutinhead{M31N 2002-08a}
2490.523 & R & 17.05 & 0.10 \cr
2504.452 & R & 17.20 & 0.10 \cr
2512.360 & R & 18.20 & 0.20 \cr
2513.344 & R & 17.83 & 0.09 \cr
2516.506 & R & 18.10 & 0.10 \cr
2517.368 & R & 18.60 & 0.20 \cr
2517.560 & R & 18.40 & 0.15 \cr
2520.452 & R & 18.40 & 0.20 \cr
\enddata
\end{deluxetable}

\end{document}